\documentclass[preprint]{aastex}
\usepackage{array, wrapfig}
\usepackage{graphicx} 
\usepackage{natbib}
\usepackage{gensymb}
\usepackage{amsmath}
\usepackage{color}
\usepackage{longtable}

\received{}
\revised{}
\accepted{}
\submitjournal{ApJ}

\shorttitle{Flare-Imminent {\it vs.} Flare-Quiet Corona through Chromosphere I: AARPs}
\shortauthors{Dissauer et al.}

\begin{document}

\title{Properties of Flare-Imminent versus Flare-Quiet Active Regions from the Chromosphere
through the Corona I: Introduction of the AIA Active Region Patches (AARPs)}

\correspondingauthor{Karin Dissauer}
\correspondingauthor{K.D. Leka}
\email{dissauer@nwra.com; leka@nwra.com}

\author[0000-0001-5661-9759]{Karin Dissauer}
\affiliation{NorthWest Research Associates, 3380 Mitchell Lane, Boulder, CO 80301 USA}

\author[0000-0003-0026-931X]{K.D. Leka}
\affiliation{NorthWest Research Associates, 3380 Mitchell Lane, Boulder, CO 80301 USA}
\affiliation{Institute for Space-Earth Environmental Research, Nagoya University, \\
Furo-cho Chikusa-ku, Nagoya, Aichi 464-8601 JAPAN}
\email{leka@nwra.com}

\author[0000-0002-7709-723X]{Eric L. Wagner}
\affiliation{NorthWest Research Associates, 3380 Mitchell Lane, Boulder, CO 80301 USA}

\begin{abstract}

We begin here a series of papers examining the chromospheric and coronal
properties of solar active regions.  This first paper
describes an extensive dataset of images from the Atmospheric Imaging
Assembly on the Solar Dynamics Observatory curated for large-sample
analysis of this topic.  Based on (and constructed to coordinate with)
the ``Active Region Patches'' as identified by the pipeline data analysis
system for the Helioseismic and Magnetic Imager (HMI) on the same mission,
the ``HARPs''), the
``AIA Active Region Patches'' (AARPs), described herein, comprise 
an unbiased multi-wavelength set
of FITS files downsampled spatially only by way of HARP-centered patch
extractions (full spatial sampling is retained), and downsampled in the
temporal domain but still able to describe both short-lived kinematics
and longer-term trends.  The AARPs database enables physics-informed
parametrization and analysis using Nonparametric Discriminant Analysis in
Paper II of this series, and is validated for analysis using Differential
Emission Measure techniques.  The AARP dataset presently covers mid-2010
through December 2018, is $\approx$9TB in size, and available through
the Solar Data Analysis Center \citep{aarp_data}.

\end{abstract}


\section{Introduction}
\label{sec:intro}

Coronal magnetic topology, energetics, and dynamics are all 
believed to play key roles in triggering, powering, enabling
energetic events.  However, large-sample studies of active regions that search for clues
as to solar energetic event productivity have long focused on
characterizing photospheric active region complexity through
the analysis of continuum-image and magnetic field data, and the
subsequent association of resulting descriptors with flare activity
\citep{flareprediction,zirinliggett87,McIntosh1990,BornmannShaw1994,McAteer_etal_2005,
dfa3,Kontogiannis_etal_2019,nci_daffs,Al-Ghraibah_etal_2015,Korsos_etal_2014}.
In addition to descriptors based on the morphology of white-light
images and the spatial distribution and character of the magnetic fields, such as 
the Zurich sunspot classification system, photospheric
analysis of solar active region flare productivity has also included 
plasma velocity and helicity patterns
\citep{Welsch_etal_2009,Park_etal_2018,Park_helicity_2},
wavelet and fractal analysis of photospheric images 
\citep{Abramenko2005b,McAteer_etal_2005,Georgoulis2012,Al-Ghraibah_etal_2015},
and inferred sub-surface plasma flows \citep{Komm_etal_2011a,Braun2016}.

The focus on photospheric magnetic fields and the drivers of their
evolution makes physical sense, as these are ultimately the source of 
energy to power solar energetic events.  The physical interpretation of 
the state of the photosphere for flare-productive active regions includes
highly non-potential magnetic fields that indicate stored magnetic 
energy, strong electric current 
systems and spatial gradients that provide pathways for magnetic 
reconnection, and emerging flux episodes that can destabilize the system 
\citep[e.g.][]{zirintanaka73,Krall_etal_1982,hagyardetal84,CPR,wangetal96,smithetal96}, 
\citep[see also][and references therein]{params,dfa3}.
These quantitatively interpretable characteristics provide the physics-based
insight that can then guide or constrain numerical modeling and further understanding
\citep[recent examples include][]{Threlfall_etal_2017}.   We intentionally
cite some of the originating literature to highlight these
physics-inspired investigations that led to {\it e.g.}, the 
Space-Weather HMI Active Region Patch \citep[``SHARP''][]{hmi_sharps} {\it parameters}
currently published as meta-data physical summaries of solar magnetic
complexes as a data product from the Solar Dynamics Observatory \citep[SDO][]{sdo}
Helioseisemic and Magnetic Imager \citep[HMI][]{hmi,hmi_cal,hmi_invert,hmi_pipe}.

In this series of papers we begin to do the same, quantitatively,
for the chromosphere, transition region, and corona.  The upper layers of the atmosphere
mirror, reflect, and react to the photospheric drivers.  
Case-studies of the pre-event chromosphere, transition region and corona have shown
evidence of specific pre-event energization and kinematics,
starting with enhanced Hydrogen Balmer-series ``H$\alpha$'' emission in flare-imminent active
regions \citep{flareprediction,BearAlerts1991},  an increase
in chromospheric non-thermal velocities and high blueshifts
\citep{Cho_etal_2016,Harra_etal_2013, Woods_etal_2017,Seki_etal_2017},
very localized chromospheric heating \citep{Li_etal_2005,Bamba_etal_2014}.
Coronal brightness and morphological signals include the energization and
increased dynamic behavior of EUV structures \citep[``crinkles'';][and
references therein]{SterlingMoore2001b,Joshi_etal_2011,
SterlingMooreFreeland2011,ImadaBambaKusano2014}, and coronal dimmings
\citep{ImadaBambaKusano2014,Zhang_etal_2017,QiuCheng2017} in the hours
prior to energetic events.

Large-sample studies of the corona that relate its physical state
to energetic-event productivity have generally focused on modeling the 
coronal magnetic field and observationally inferring relevant topological characteristics,
often for coronal mass ejections \citep[{\it e.g.} ][]{Barnes2007,dfa2,GeorgoulisRust2007,Kusano_etal_2020}
\citep[although see][regarding filament dynamics and eruptivity]{Aggarwal_etal_2018}.

Thus far, investigations using the large-sample data available from 
the SDO Atmospheric Imaging Assembly \citep[AIA][]{aia_Lemen} 
in the context of energetic events have been
carried out with a primary goal of forecasting, rather than physical understanding.
To achieve this expressed objective, machine-learning tools
have been invoked \citep{Nishizuka_etal_2017,Jonas_etal_2018,Alipour_etal_2019}, 
although these routinely subject the AIA data to significant degradation
by spatial binning, which hampers
the ability to discern any small-scale processes.
The outcomes have focused on skill scores rather than physical
interpretation, as ``interpretable machine-learning'' approaches have 
yet to be widely implemented in this context.
In other words, on many levels, information about the physics
of the corona has been lost or not pursued.

We focus here on the need to move away from case studies and
toward the scientific objective of quantitatively characterizing 
the behavior of the chromosphere
and corona using physically-meaningful analysis.  We ask parallel
questions about the chromosphere and corona as were posed in previous works 
that centered on the photosphere
\citep{params,dfa,dfa2,dfa3}: what are the broad physical differences
between ``flare-imminent'' and ``not-flare-imminent'' active regions?
For the photosphere, we focused on descriptors of the vector magnetic
field and parametrizations that quantify, for example, the free magnetic energy
available for energetic events.
For the present objectives, we focus on both high-frequency
kinematic analysis indicative of small-scale reconnection activity,
and the hours-long trends in the brightness distributions indicative
of growth, decay, or changing temperature distribution, or 
increasing or decreasing levels of small-scale activity.
We design the AARP database to address these objectives 
with a statistically-significant sample.
As such, while we are {\it not} actively focusing on flare ``prediction'' 
{\it per se}, the results of these investigations may inform 
those activities eventually, we propose that the AARP database is 
designed such that it is appropriate for other, very different, scientific objectives, as well.

In this first paper of the series, we describe the data preparation for, and full description of, 
the AIA Active Region Patch extractions 
\citep[AARPs;][see Section\,\ref{sec:aarps}]{aarp_data}.
We present global, cycle-scale trends across wavelengths 
in Section\,\ref{sec:analysis}, plus some additional analysis 
aimed to motivate community use of the AARP dataset.
We leave the full discussion of physics-informed parametrization and results from Nonparametric 
Discriminant Analysis to \citet[Paper II; ][]{nci_aia}, and of a statistical analysis
of the temperature and density of flare-imminent regions using Differential Emission Measure 
tools to \citet[Paper III;][]{nci_dem}.

\section{The AIA Active Region Patch (AARP) Database}
\label{sec:aarps}

\begin{figure}[t]
\centerline{
\includegraphics[width=1.0\textwidth,clip, trim = 0mm 0mm 0mm 0mm, angle=0]{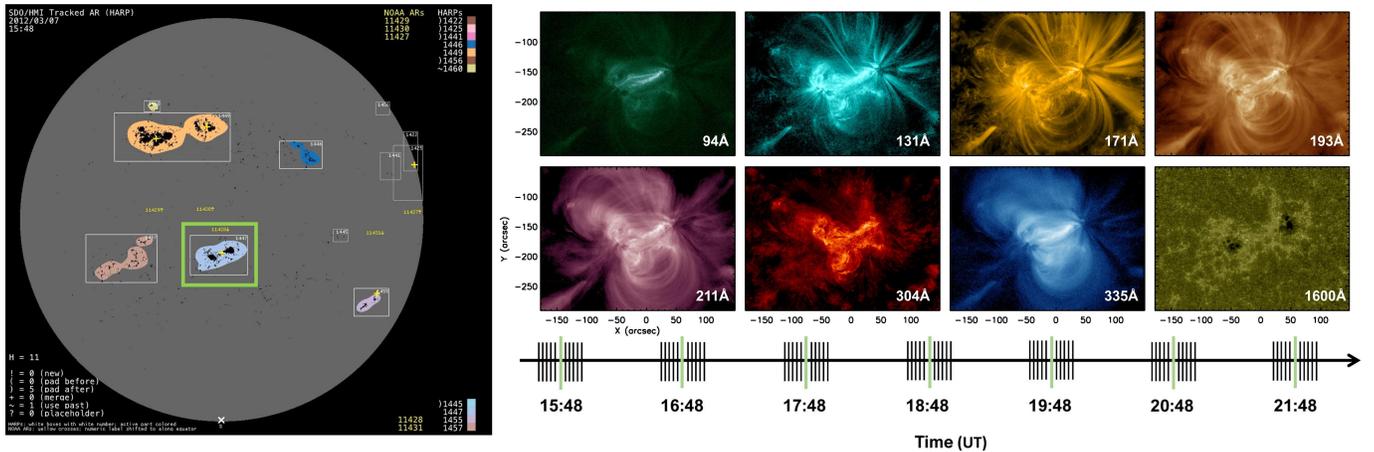}}
\caption{Overview of the AARP extraction and down-sampling in both spatial and temporal domains, 
for HARP/AARP 1447 on March 07, 2012.
The AARP field-of-view is based on the HARP box (highlighted by the green boundary in the left panel; 
see Section~\ref{sec:aia_spatialdownselect}). In the temporal domain, 11 images (right panel) per hour 
(see time line at the bottom) over the course of 1/4 of each day are extracted for 
8 different (E)UV wavelengths (all but 1700\AA; panels in upper/right).}
\label{fig:aarp_extraction}
\end{figure}

In this section we describe the AIA Active Region Patch (AARP) Database
and its construction (see Figure~\ref{fig:aarp_extraction}).   The data cubes were designed to make available
a statistically significant sample of solar active region coronal and chromospheric imaging data
with a database of tenable size (noting that the full-disk full-cadence 
AIA dataset for our target period is over 25PB).  The resulting cubes are the basis for subsequent
analysis \citep[e.g.][]{nci_daffs} and are available to the community \citep{aarp_data}.

In the present incarnation, we specifically select the cadence and
duration of the cubes to match those we have produced for photospheric
active region investigations using the vector-field data from 
HMI \citep{nci_daffs}; future studies may require
different such particulars.  The infrastructure to produce the present 
dataset as well as others with different such options 
is available \citep{aarp_data}, although it presently
relies on specific data access and architecture (see Appendix~\ref{sec:appendix_drms},~\ref{sec:appendix_cutout}).
Not everyone in the community has access to the hardware and software
resources required to produce the AARP dataset; we hope it will be 
of use to the community as a curated, ready-for-analysis resource.

The HMI Active Region Patch \citep[``HARP''][]{hmi_pipe,hmi_sharps} definitions
are used for the active-region identification system, and we intentionally include all HARPs.  We do 
not down-select at the active-region
level, either for size or activity.  HARP designations are based on 
concentrations of magnetic activity, and the low-activity, small concentrations
statistically dominate.  As such, we avoid imposing a bias for later analysis,
although any comparison of (for example) active {\it vs.} less-active regions must 
statistically account for the uneven sample sizes.  

\subsection{AIA Data}
\label{sec:aia_data}

The AARP database comprises coronal- and chromospheric
time-series patches that are extracted from the full-cadence
full-size data of the Atmospheric Imaging Assembly
\citep[AIA;][]{aia_Lemen} on board NASA's Solar Dynamics Observatory
\citep[\textit{SDO};][]{sdo}.  AIA obtains $4096^2$-pixel full-disk
coronal and chromospheric images at $1.5^{\prime\prime}$ spatial resolution, sampled at $0.6^{\prime\prime}$ through 
multiple extreme-ultraviolet filter bands at [94, 131, 171, 193, 211, 304, 335]\,\AA\ with a 12\,s 
cadence, and full-disk photospheric images at ultraviolet [1600, 1700]\,\AA\ with a 24\,s cadence.  
We embarked on this project to avoid copying {\it all} of the data,
instead focusing on active-regions specifically for the first
``down-selection'' operation for the final database.  However, that 
still required handling a significant amount of AIA data.

NWRA employs the Joint Science Operations Center (``JSOC'') ``NetDRMS''
(Network Data Record Management System) / ``RemoteSUMS''(Storage Unit
Management System) system to be able to essentially mimic the availability
of {\it SDO} (and other) data in a manner directly analogous to the
host institution, Stanford University.  Large data transfers are handled
through the ``JSOC Mirroring Daemon'' (see Appendix~\ref{sec:appendix_drms}).
The AIA meta-data reside in the {\tt aia.lev1\_euv\_12s} and 
{\tt aia.lev1\_uv\_24s} series, while the images
reside in the {\tt aia.lev1} series within the NetDRMS/RemoteSUMS system (the {\tt aia.lev1} series
is largely transparent for data transfers through, {\it e.g.}, the JSOC
``LookData'' facility).  Details of setting up automatic targeted Java Mirroring Daemon (JMD) transfers
and performing the extractions are provided in Appendix~\ref{sec:appendix_drms}.
For small numbers of HARPs or customized extractions,
the {\tt SolarSoft} {\tt ssw\_cutout\_service.pro} \citep{solarsoft} is recommended to be used,
however as described in Appendix~\ref{sec:appendix_cutout}, it was not 
the correct tool by which to develop the AARP database.

\subsection{Down-Select in the Spatial Domain: Extracting the Active Region Patches}
\label{sec:aia_spatialdownselect}

This dataset is created to address the almost untenable requirements of large-sample
coronal studies. 
The ``AIA Active Region patches'' (AARPs) are first and foremost a down-selection
in the spatial domain, providing sub-area extractions
from the full-disk AIA image data that rely on the HMI Active Region Patches 
\citep[HARPs;][]{hmi_pipe} metadata ({\tt hmi.Mharp\_720s} series) for target
selection, pointing, and field-of-view (FOV) definition.

\begin{figure}[t]
\centerline{
\includegraphics[width=0.565\textwidth,clip,trim = 7mm 2mm 21mm 12mm, angle=0]{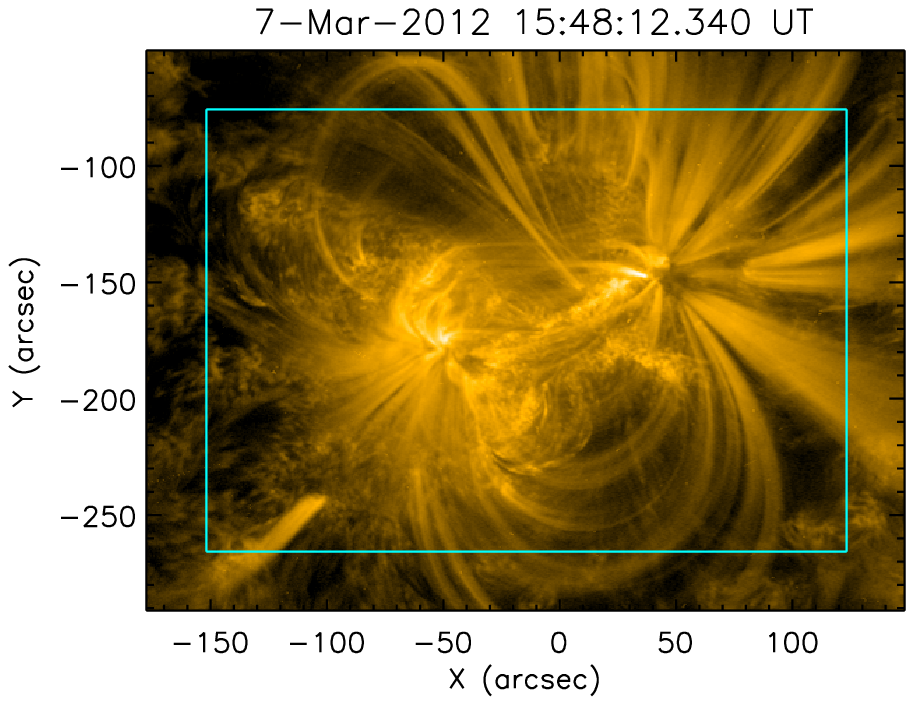}
\includegraphics[width=0.435\textwidth,clip,trim = 20mm 2mm 30mm 12mm, angle=0]{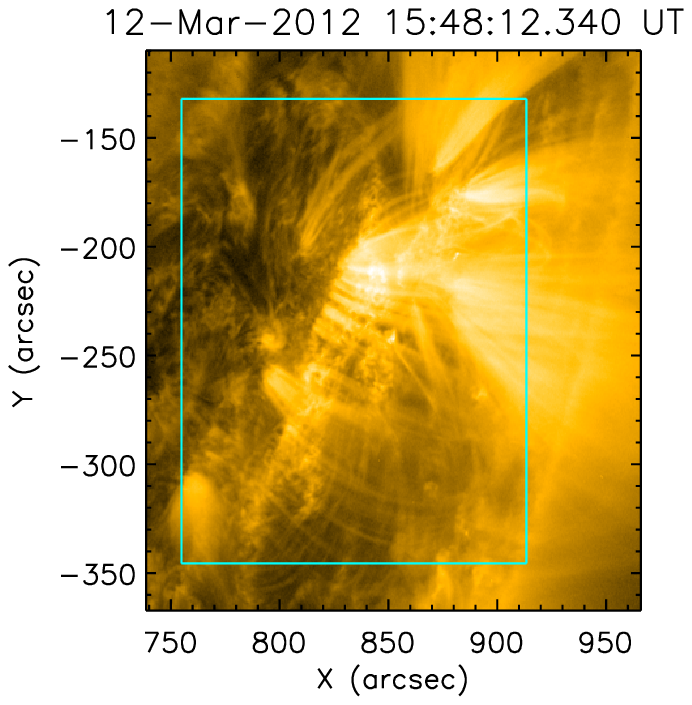}}
\caption{Two examples of the same AARP in 171\AA~illustrating the field-of-view expansions
as a function of location on the disk and HARP dimensions.
Symmetrical (left) extensions are used near disk center, and asymmetric extensions to 
include coronal loops are employed as regions approach the limbs, {\it e.g.} West 
(right) limbs. 
The cyan box marks the original HARP field-of-view. 
Left: AARP\,\#1447 (NOAA\,AR\,11428) 2012.03.07 15:48~UT; Right: AARP\,\#1447 (NOAA\,AR\,11428), 
2012.03.12 15:48~UT.  
}
\label{fig:overview_aarp}
\end{figure}

Because coronal structures extend in height from the solar surface, the 
extraction box is modified from the original HARP FOV to accommodate this projection
effect both when a region is viewed face-on but also as it
is viewed away from disk center (see Figure~\ref{fig:overview_aarp}).
Hence, we expand the original
HARP FOV by 20\% in both x-, y- directions, limited to a maximum expansion of 25\arcsec~to avoid 
unwieldy increases in already-large HARPs. Additionally, we
augment with an expansion towards the East or the West $x_{\text{ex}}$ depending on the position of the
HARP center with respect to the central meridian in the form of
\begin{equation}
    x_{\text{ex}}=x_{\text{cen}}[^\circ]/90 ^\circ \cdot 50\arcsec \, ,
\end{equation}
where $x_{\text{cen}}$ is the center of the HARP in degrees. This
expansion is limited to a maximum extension of 50\arcsec~and is done
to accommodate the additional extension of loops viewed from the side
as a region approaches the solar limb. Figure~\ref{fig:overview_aarp}
shows two examples, a disk-center and a near-limb target, comparing the original 
HARP (cyan box) and resulting AARP fields of view.  
While loop tops originating from the target active region may still be cut off,
the primary science this dataset is designed for concerns relative (vs. absolute) changes in AR behavior, 
and since all regions are treated consistently, this should impose no statistical concerns.
In addition, automatically defining an optimal FOV for each individual active region is challenged by 
(1) AARPs frequently being in close proximity during periods of high activity, and (2) an optically-thin corona, 
such that there is arguably no optimal way to automatically disentangle all visible loops, their source
active regions, and still ensure their tops are always included and correctly assigned.

\begin{figure}[t]
\centerline{
\includegraphics[width=0.33\textwidth,clip, trim = 5mm 14mm 7mm 3mm, angle=0]{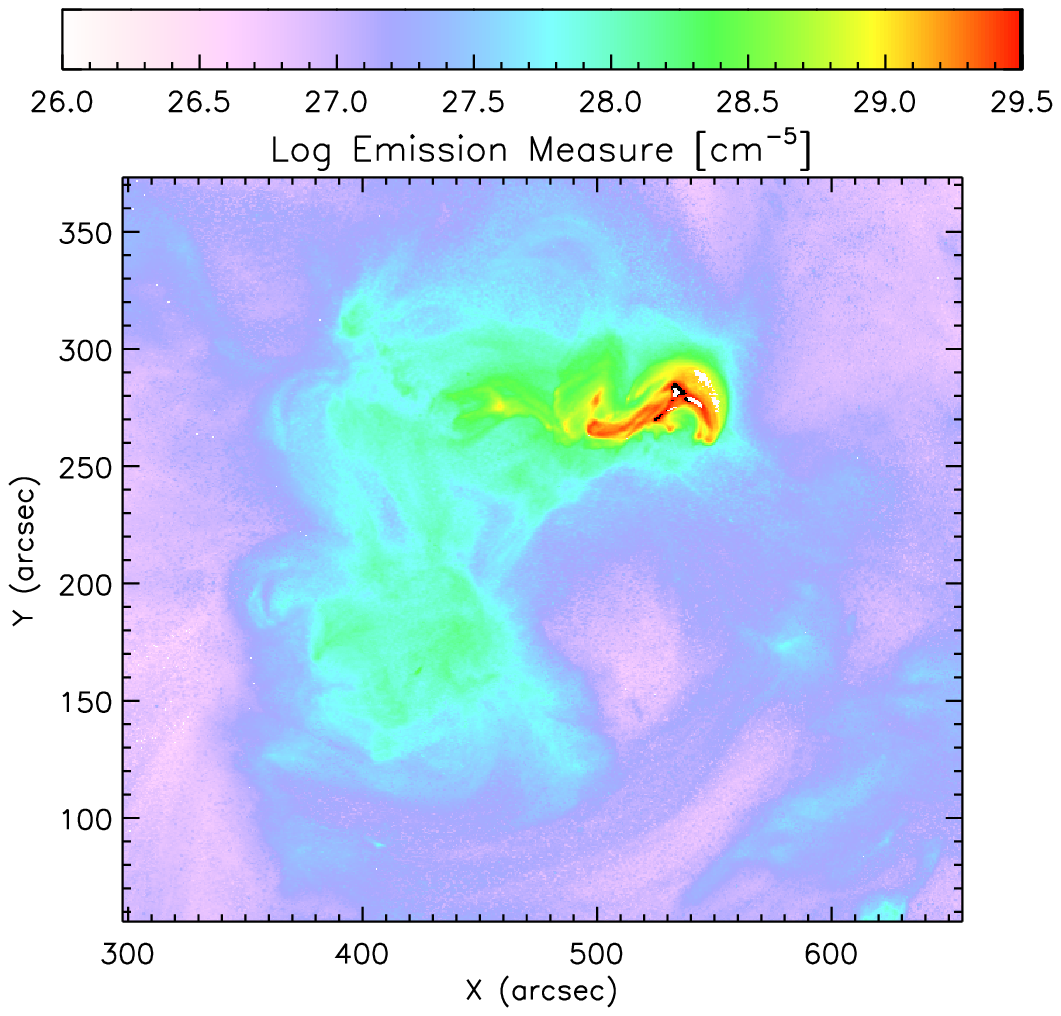}
\includegraphics[width=0.33\textwidth,clip, trim = 5mm 14mm 7mm 3mm, angle=0]{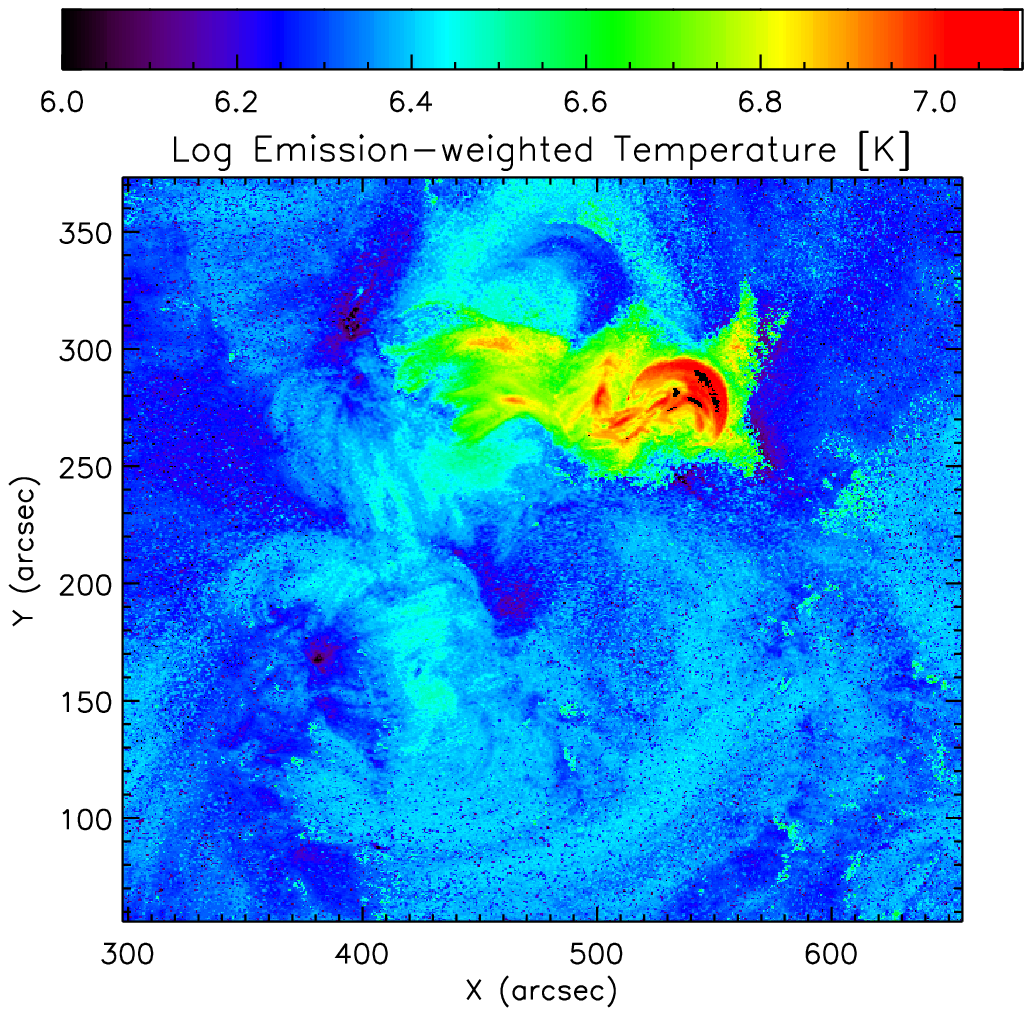}
\includegraphics[width=0.33\textwidth,clip, trim = 5mm 14mm 7mm 3mm, angle=0]{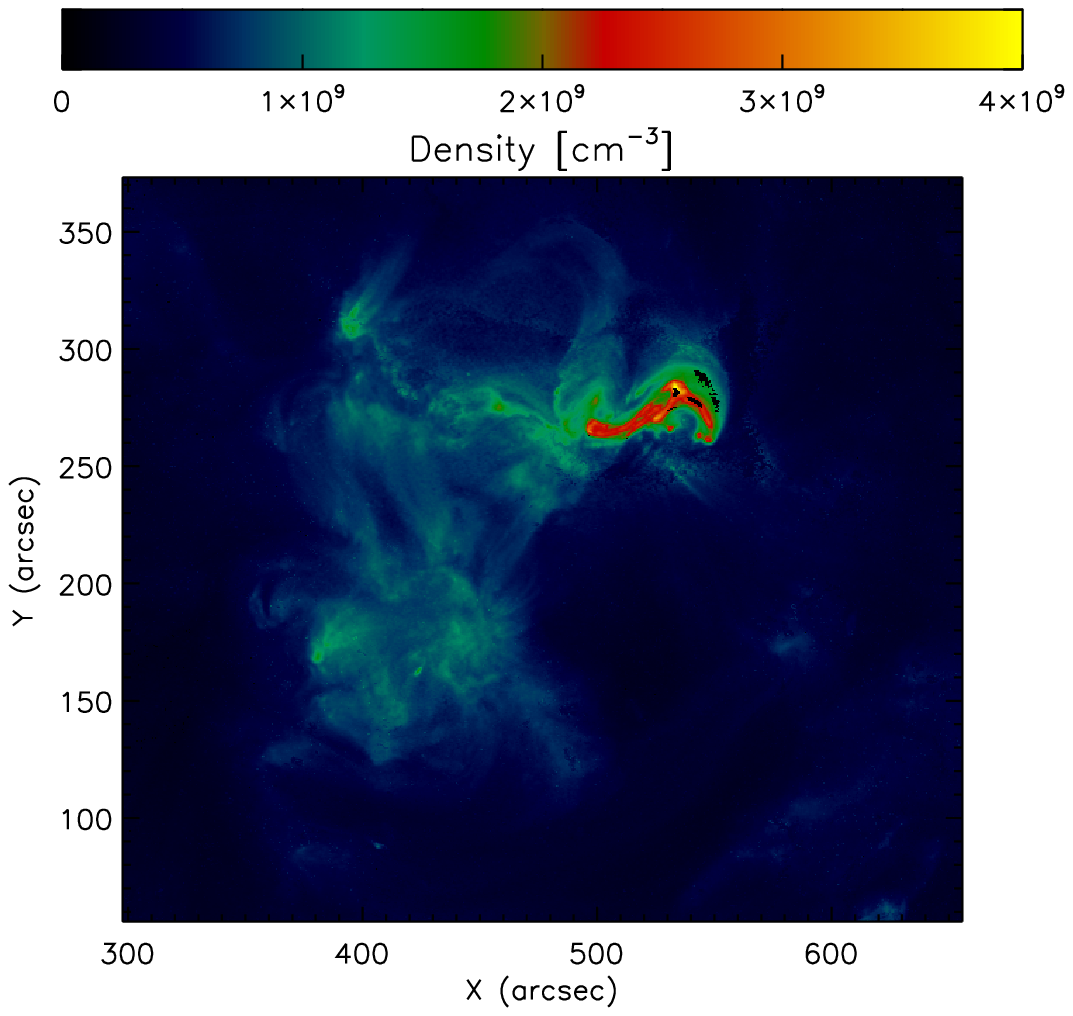}}
\caption{Analysis using Differential Emission Measure \citep{Cheung_etal_2015} for HARP/AARP 3894 
(NOAA\,ARs 12017 \& 12018) on March 29, 2014 at 17:42:01 UT.
Left: emission measure map (log-scale), Middle: emission-weighted temperature map (log-scale), 
Right: density map.
}
\label{fig:dem_aarp}
\end{figure}

All HARPs are included regardless of flaring activity levels, size of
active regions, or observing angle.  The originating HARP bounding-box 
information is used for the central-time (the ``XX:48'') image targets. 
The extracted higher-cadence 11-image timeseries
is co-aligned between the (nearly co-temporal) wavelengths 
and the central-time image using differential rotation to allow for
Differential Emission Measure (DEM) analysis (Figure~\ref{fig:dem_aarp};
see also Appendix~\ref{sec:appendix_DEM}) but no further tracking is needed.

\subsection{Down-selecting in the Temporal Domain}
\label{sec:aia_temporaldownselect}

To reduce the data load while preserving the scientific requirements of the research 
performed in Paper II and further projects, we down-sample the data in the temporal domain.
The native cadence for AIA EUV images is 12\,s, and for the UV images
it is 24\,s.  The AIA Automatic Exposure Control (AEC) system is an
observation mode designed to avoid image saturation during solar flares.
When the AEC system is invoked, the AEC-regulated images for the EUV data
are inter-leaved with normal-exposure (possibly saturated) images,
reducing the effective cadence for consistent-exposure EUV data to 24s.
Native images in data numbers (``DN'') or counts can be normalized by the exposure
time to obtain consistent per-pixel count rate-based images. For reasons described
below, however, we decided to avoid the AEC-regulated images and sample
the AIA images at a 72\,s cadence.

\begin{figure}[t!]
\centerline{
\includegraphics[width=0.33\textwidth, clip, trim = 2mm 5mm 2mm 10mm]{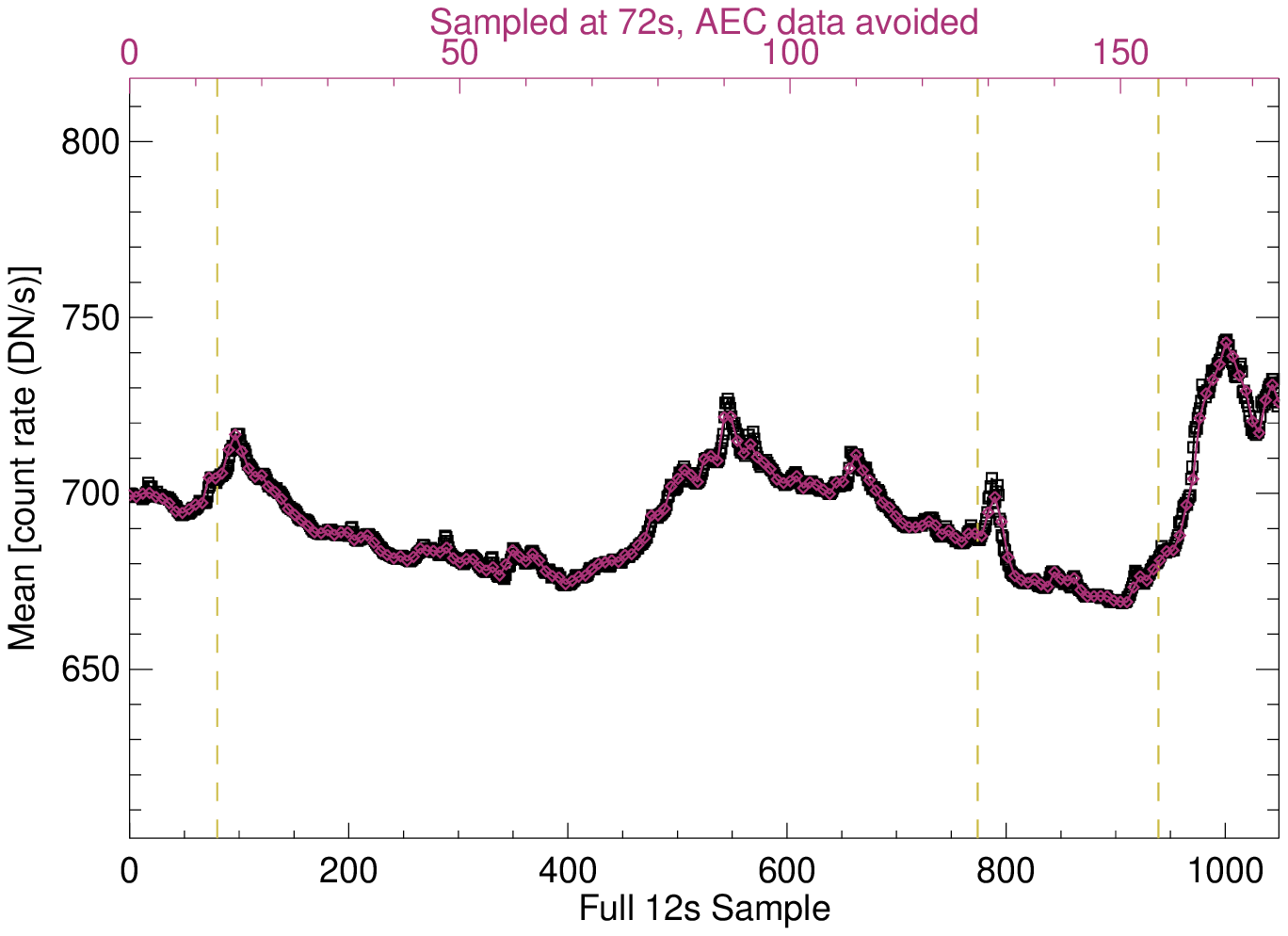}
\includegraphics[width=0.33\textwidth, clip, trim = 2mm 5mm 2mm 10mm]{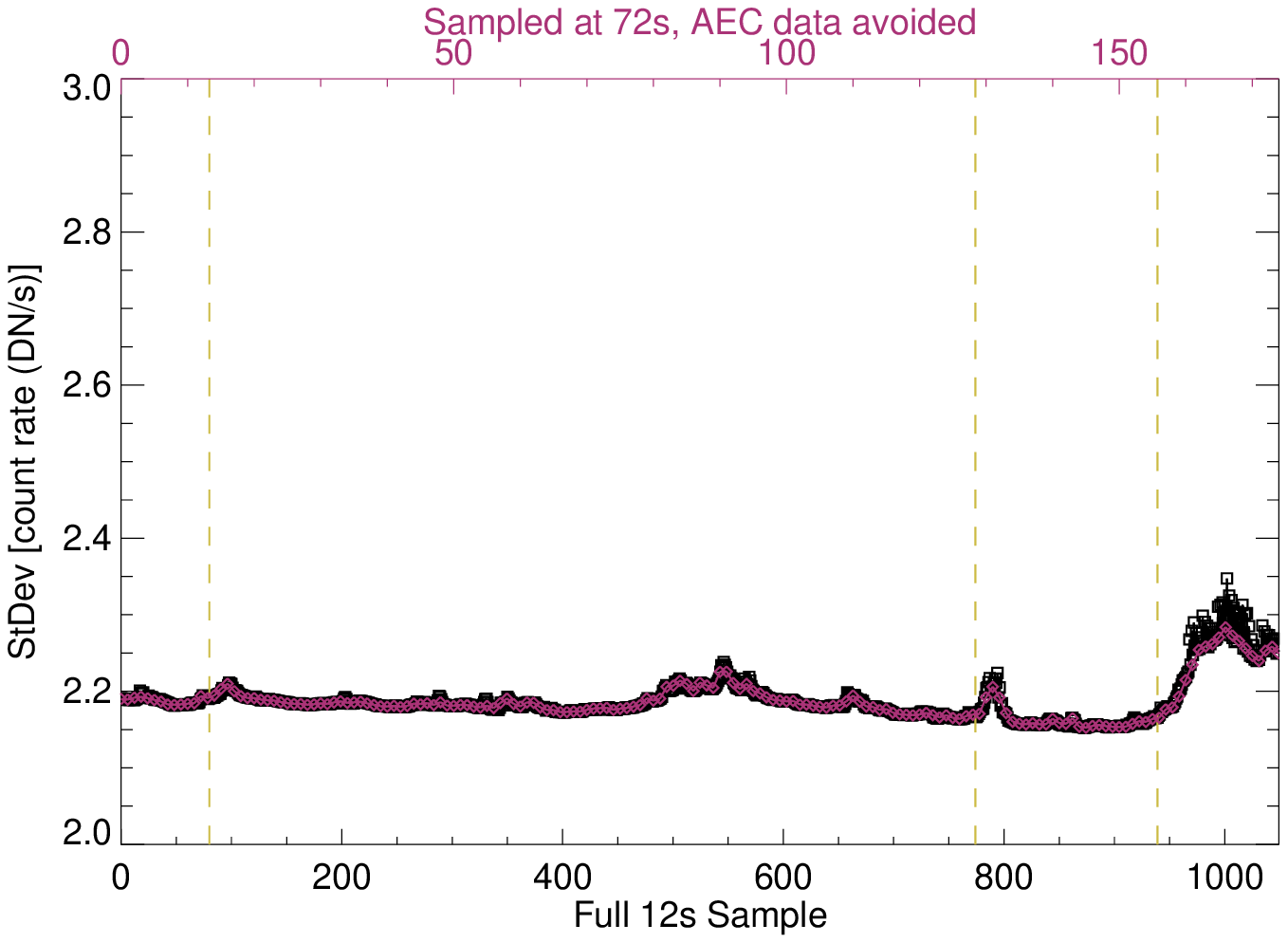}
\includegraphics[width=0.33\textwidth, clip, trim = 10mm 5mm 2mm 3mm]{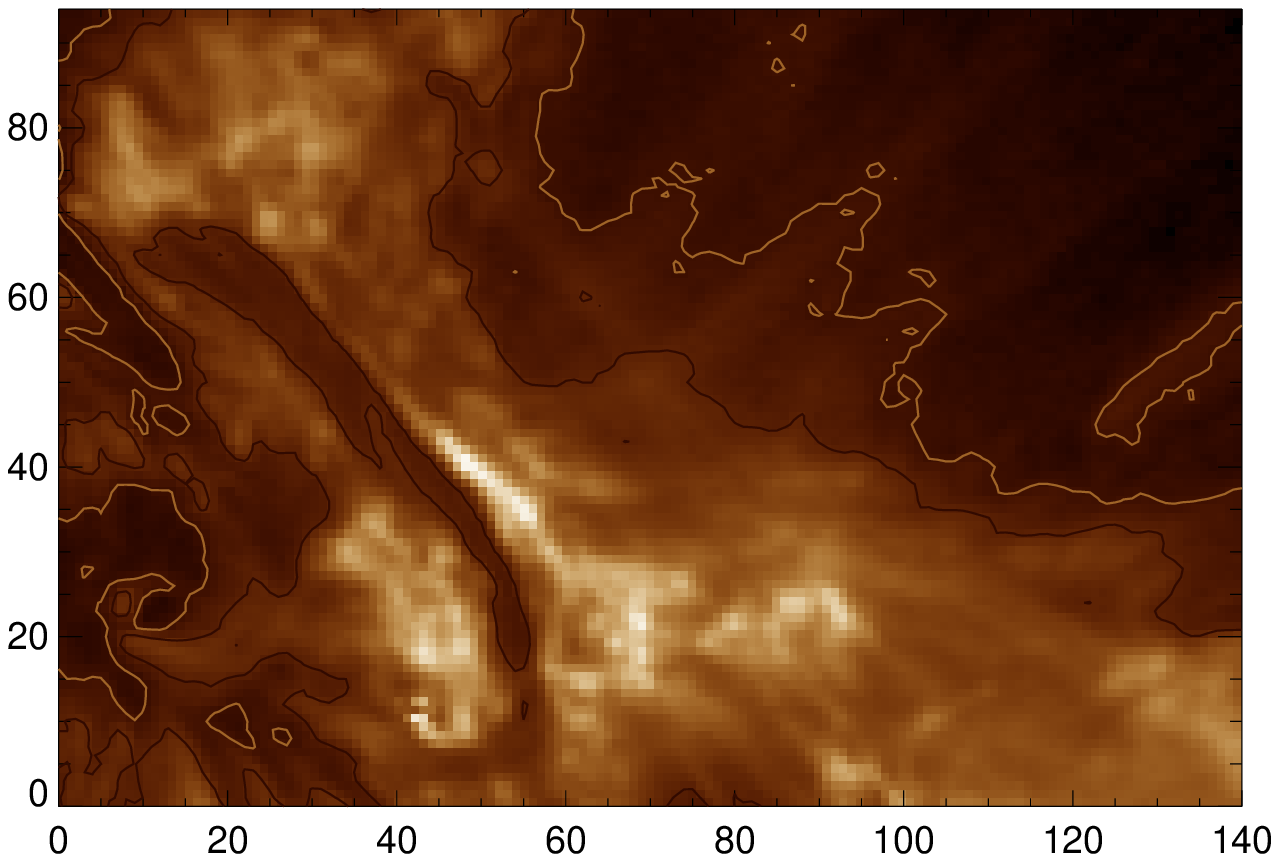}}
\centerline{
\includegraphics[width=0.33\textwidth, clip, trim = 2mm 5mm 2mm 10mm]{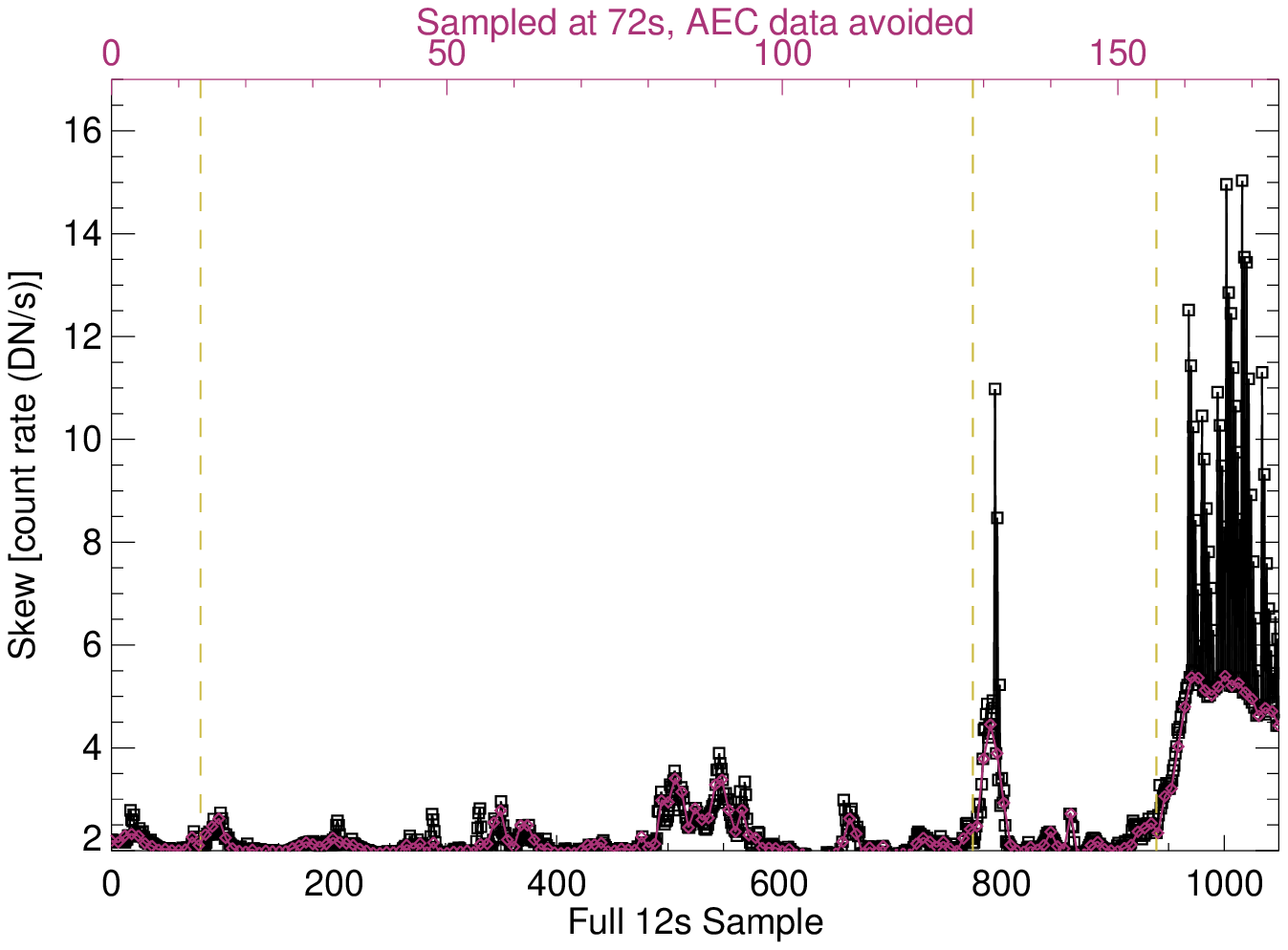}
\includegraphics[width=0.33\textwidth, clip, trim = 2mm 5mm 2mm 10mm]{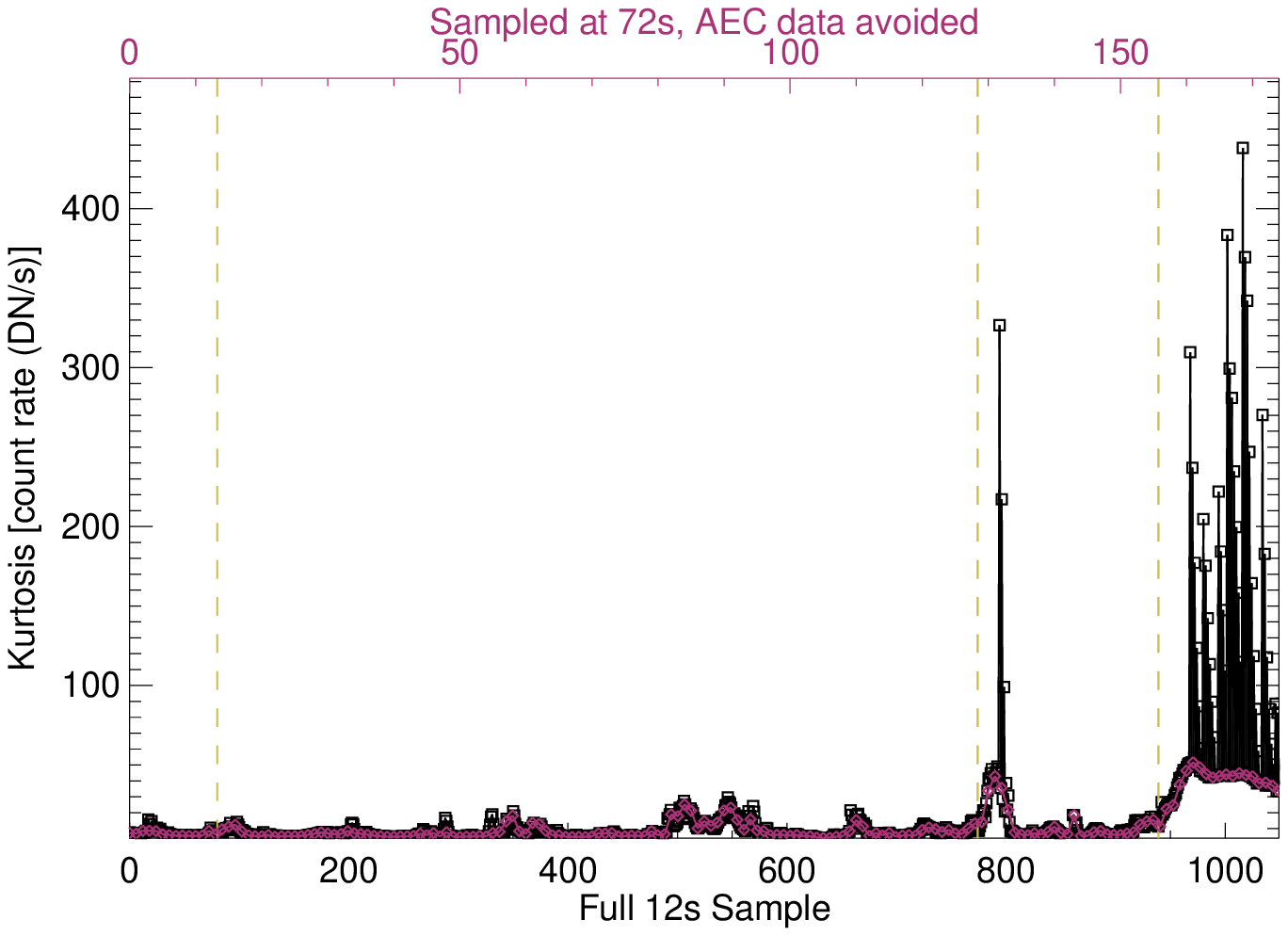}
\includegraphics[width=0.33\textwidth, clip, trim = 10mm 5mm 2mm 3mm]{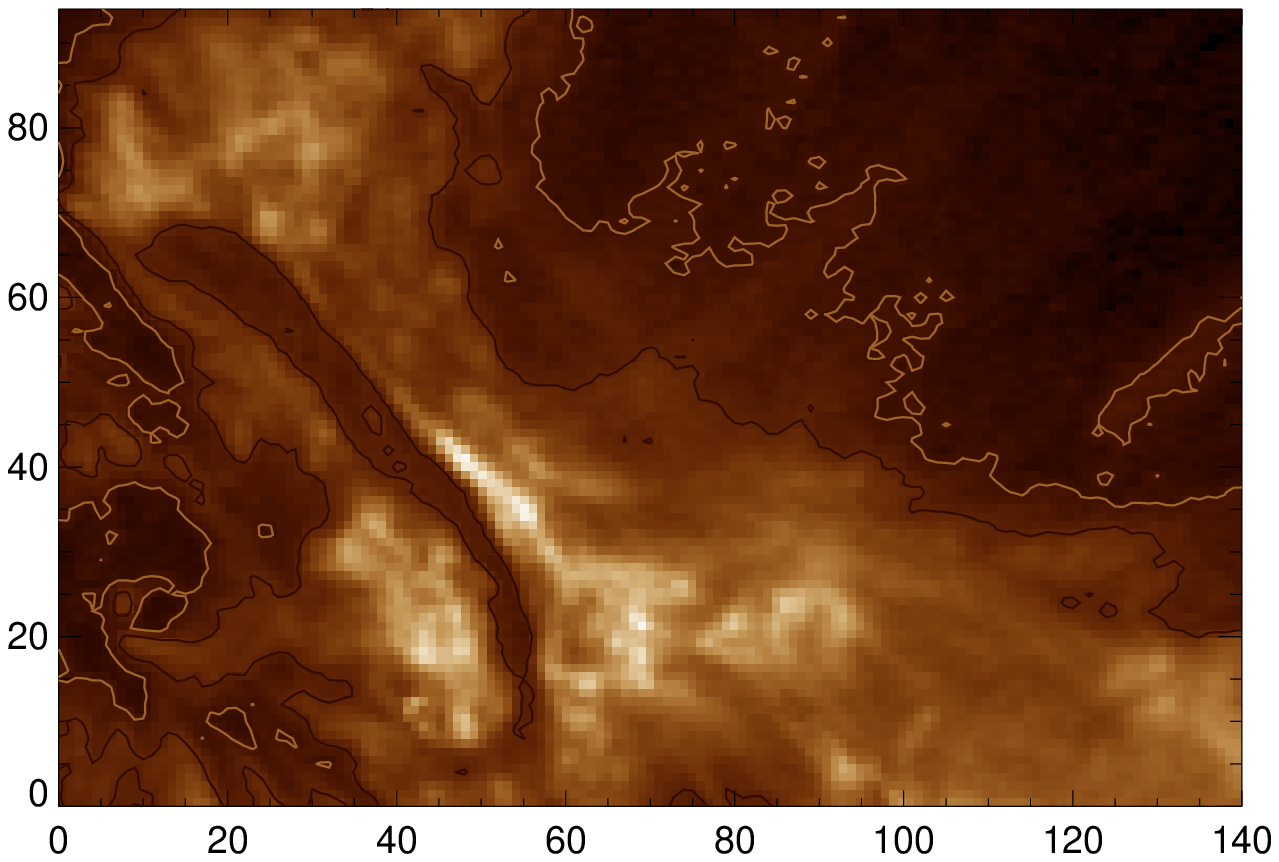}}
\caption{Example parameter curves from AARP~\#750
(NOAA AR~\#11261) for 193\AA\ on 2011.08.03 00:00UT~--~03:30~UT; 
the start of a sub-flare at 00:18~UT, a C1.2 at 02:35~UT and the 
M1.1 flare at 03:08~UT are indicated by dashed vertical lines.
The bottom x-axis is an image index for the original 12\,s sampling (black),
the top x-axis are the indicies for the 72\,s sampling (red) used for the AARPs
that naturally avoided the AEC-triggered images.  All data were exposure-normalized.
Left four panels show the four moments of the images, as indicated.
Right two panels show a small sub-area of the full AARP field of view that
includes a small filament and nearby loop-bottoms (top/right of the images), but avoids
the M1.1 flare-emission area, for two consecutive 
12\,s-sampled images, {\bf Top}:
normal 2\,s exposure, {\bf Bottom}: 0.43\,s exposure, when AEC-mode was invoked.
Light/Dark contours indicate the [2, 3]\,$\sigma$ 
signal/noise levels, with the noise level determined by the most-probable value of low-signal 
parts of the image.}
\vspace{-0.25cm}
\label{fig:aiacurves}
\hrulefill
\vspace{-0.5cm}
\end{figure}

The image parametrization for the statistical analysis
in Paper II and later work uses moment analysis to describe
the coronal and chromospheric behavior.  
During flare times, even for small flares, the higher moments
jump dramatically -- as expected in the presence of localized
yet high-magnitude brightness changes (Figure ~\ref{fig:aiacurves}, left 
four panels).  Our scientific focus is not, generally, the flaring plasma 
itself but the broad active region behavior.  Avoiding the AEC
images allows for times of flares to be included with less dramatic 
impact to the moment analysis.  Additionally, the shorter exposures
during flaring times lead to lower signal/noise ratios (SNR) for the
larger active region areas (Figure ~\ref{fig:aiacurves}, right panels).
Since these regions are our focus, ensuring a consistent SNR is key.
Finally, given the format of the AARP FITS files (see Appendix~\ref{sec:appendix_header}),
avoiding the AEC data allows a simplified metadata format, as the 
exposure time is consistent for a particular wavelength.

When working with running-difference images, central to our 
analysis approaches, the native cadence presents two challenges:
(1) for consistent-exposure sequential images the resulting SNR
was very low, and (2) when sequential images had varying exposure
times, the resulting variation in the images' SNR contributed to 
even higher noise in the running-difference images.  While the 
running-difference magnitudes increase as $\Delta t$, one must
also balance this increase in SNR with the ability to evaluate
and interpret the resulting images.
Experimentation indicated that the full time resolution is unnecessary
to capture the short-lived brightenings and kinematics that we wish
to examine.

A coarser 72\,s cadence for the AARP ``high-cadence'' bursts of 
images thus provides a good balance of consistent and acceptable
SNR and dataset volume reduction.  Selecting a 72\,s cadence is also a ``common
factor'' between the EUV and UV native sampling, allowing for the dataset
to sample the different atmospheric layers at the same cadence.

The evolution of the corona over multiple hours is of high
interest here.  An interval covering $\approx6$\,hr is motivated
by numerical models of pre-event evolution and trigger formation
\citep{IshiguroKusano2017,Inoue_etal_2018}, and provides sufficient
data from which to quantify overall evolutionary trends within an
active region.  Still, 72\,s sampling over 6\,hr is a large data load.
Hence we further down-select to $\approx\,13$\,min of images at the
72\,s cadence, centered hourly 15:48 -- 21:48\,UT (seven hourly
``bursts'' of images, over six hours, inclusive).  The choice of timing is driven by an
already-developed set of HMI vector-field time-series extracted data-set
\citep{nci_daffs}.  Centering the data on the ``XX:48'' times was 
necessary to avoid those HMI
data affected by calibration sequences (taken at 00:00 and 18:00\,TAI,
see \citet{hmi_pipe}).  Those studies also had a scientific 
requirement to minimize the
statistical impacts on time-series analysis that were solely due to the
spacecraft orbital velocity-induced artifacts \citep{hmi_pipe}.  Finally,
we chose 11 images covering 13\,m to effectively match the time period over which
the input spectra are averaged for the \textsf{hmi.B\_720s} series magnetograms.
In this sense, we sample the upper atmosphere on a cadence appropriate to 
its physics (reconnection, heating, flows) but carefully pair the extractions to their
arguably slower-evolving photospheric driver, at least at the HMI spatial resolution.

Thus, in summary, with the AARPs dataset the detailed behavior of the lower solar 
atmosphere is sampled at a fair cadence (72\,s) for 13\,m every hour,
repeatedly over a quarter of a day for a full 8.5\,yr (06/2010 -- 12/2018) over 
which we can query about active region trends using a large-sample approach that 
enables robust statistical analysis.

\subsection{Final Data-Cube Preparation}
\label{sec:data_final}
All AIA data are processed through the {\tt SolarSoft} \citep{solarsoft}
routine {\tt aia\_prep.pro} and a correction to account for
time-dependent degradation of the instrument was applied using the
following form:
\begin{equation}
    \frac{A_{eff}(t_{obs})}{A_{eff}(t_{0})}\left(1+p_{1}\delta t+p_{2}\delta t^{2}+p_{3}\delta t^{3} \right) \; ,
\end{equation}
where $A_{eff}(t_{obs})$ is the effective area calculated at the calibration
epoch for $t_{obs}$, $A_{eff}(t_{0})$ is the effective area at the
first calibration epoch (i.e. at launch), $p1$, $p2$, $p3$ are the
interpolation coefficients for the $t_{obs}$ epoch, and $\delta t$
is the difference between the start time of the epoch and $t_{obs}$ \citep{Barnes:2020}.

All wavelengths and imaging data centered at each hour are coaligned
and differentially rotated to the central time step (usually **:48~UT)
using {\tt drot\_map.pro}. The pre-processed output data are therefore
ready for running-difference analysis and Differential Emission Measure
(DEM) analysis (see Appendix~\ref{sec:appendix_DEM}).  Full FITS headers
are generated with wavelength, HARP number, NOAA Active Region (AR)
number (from the relevant keywords in the {\tt hmi.MHarp\_720s} series).
The seven sets of hourly image sets are saved as extensions each with
its own header recording times, number of valid images, and detailed
pointing information as well as additional keywords imported from the
{\tt hmi.MHarp\_720s} series such as {\tt LAT\_FWT, LON\_FWT} and the
World Coordinate System \citep[WCS;][]{Thompson2006} keywords for the
target 720s-based mid-time. Appendix~\ref{sec:appendix_header} contains
examples of these AARP fits headers.

The final data product thus comprises a set of eight FITS files, one
for each wavelength (by default: 94, 131, 171, 193, 211, 304, 335, and 1600 \AA),
for each HARP covering six hours inclusive (seven hourly samples 15:48 -- 21:48~UT)
of 13 minutes of data sampled at 72\,s, resulting in 11 images per hour.  
The full dataset is summarized in Table~\ref{tbl:AARPS}, its archive size being
only $\approx$ 9.5 TB, in contrast to the size of the full-disk AIA dataset for this 
period which is $\approx$25PB.  
The power of the dataset is its large, yet tenable sample size, designed to enable
unbiased statistical analysis on spatially native-resolution images.  We next present
some topics for which models or case-studies may now be tested on a statistically
significant, cycle-covering sample.

\begin{table}[h!]
\begin{center}
\caption{AARP Data Set Summary \label{tbl:AARPS}}
\begin{tabular}{|ccccc|}
\hline \hline
Date Range & HARP Range & NOAA AR Range & Total Number of Samples & Archive Size \\ \hline
06/2010 -- 12/2018 & 36 -- 7331 & 11073 -- 12731 & 256,976  & $\approx$ 9.5~TB  \\\hline
\end{tabular}
\end{center}
\end{table}

\section{Analysis \& Results}
\label{sec:analysis}

We present here a brief analysis to highlight the breadth of 
physics investigations that could be performed using the AARP database.
We do not discuss flare- or event-related studies in depth,
deferring this topic to Paper II  (although see Section~\ref{sec:dir_vs_diff}).

\subsection{Active Region Coronal Behavior with Solar Cycle}
\label{sec:ssn}

With this extensive database we can begin to examine some subtleties
of coronal behavior for active regions as a function of solar 
cycle.  Generally speaking the AARPs will not include significant
coronal hole areas, and thus will avoid that contamination and source
of confusion with regards to interpretation.  

\begin{figure}
\centerline{
\includegraphics[width=0.50\textwidth,clip, trim = 0mm 0mm 0mm 0mm, angle=0]{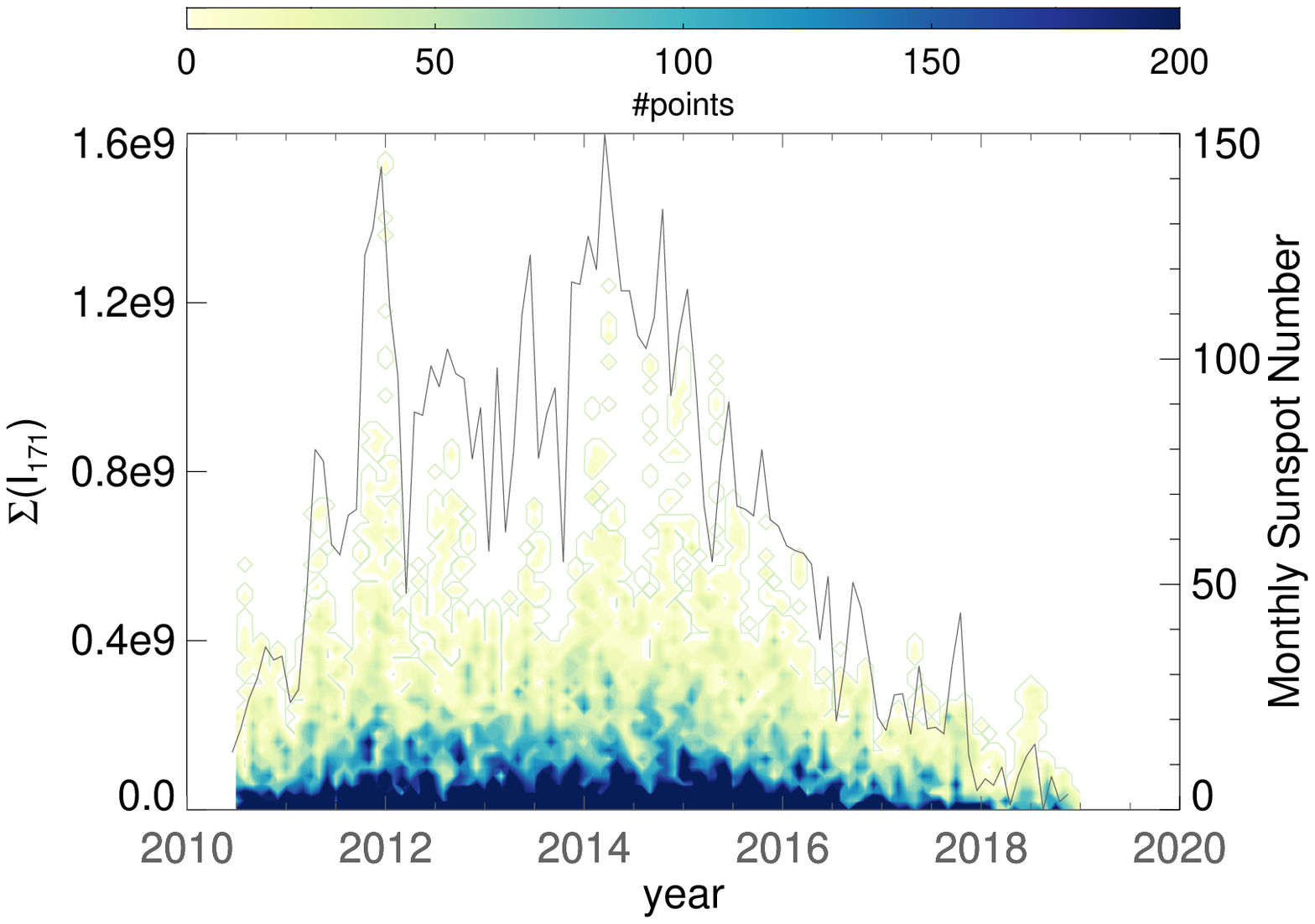}
\includegraphics[width=0.50\textwidth,clip, trim = 0mm 0mm 0mm 0mm, angle=0]{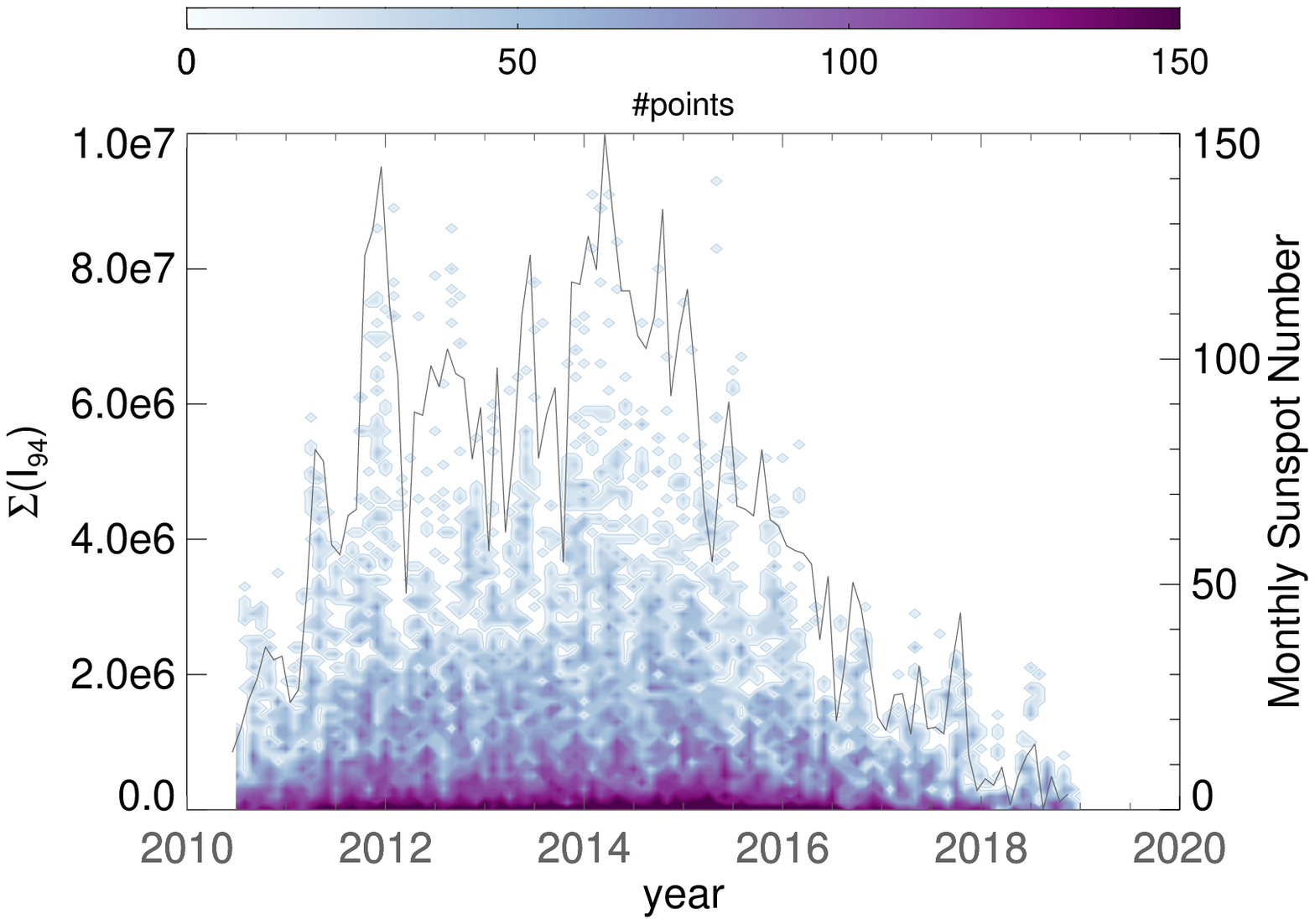}}
\centerline{
\includegraphics[width=0.50\textwidth,clip, trim = 0mm 0mm 0mm 0mm, angle=0]{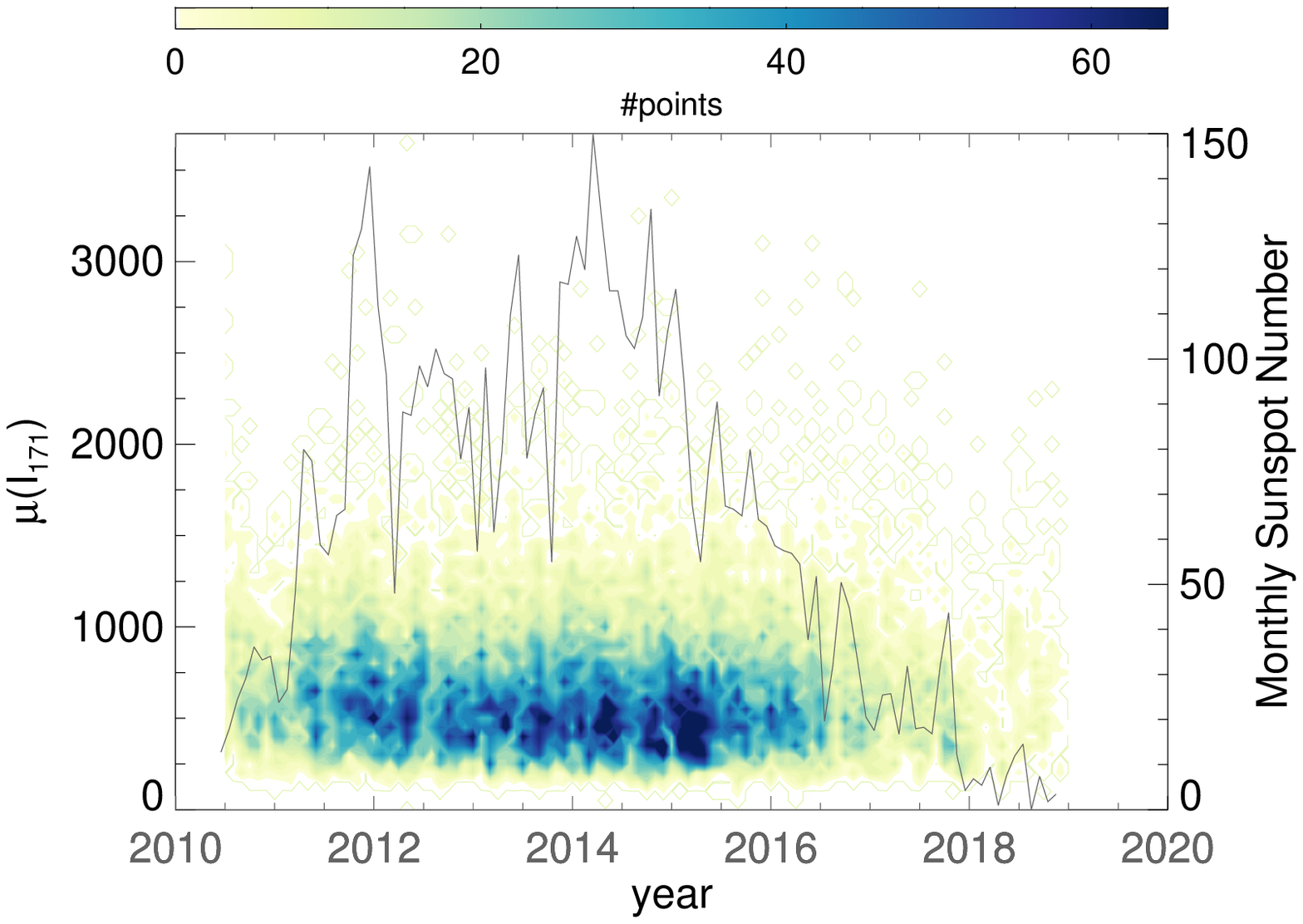}
\includegraphics[width=0.50\textwidth,clip, trim = 0mm 0mm 0mm 0mm, angle=0]{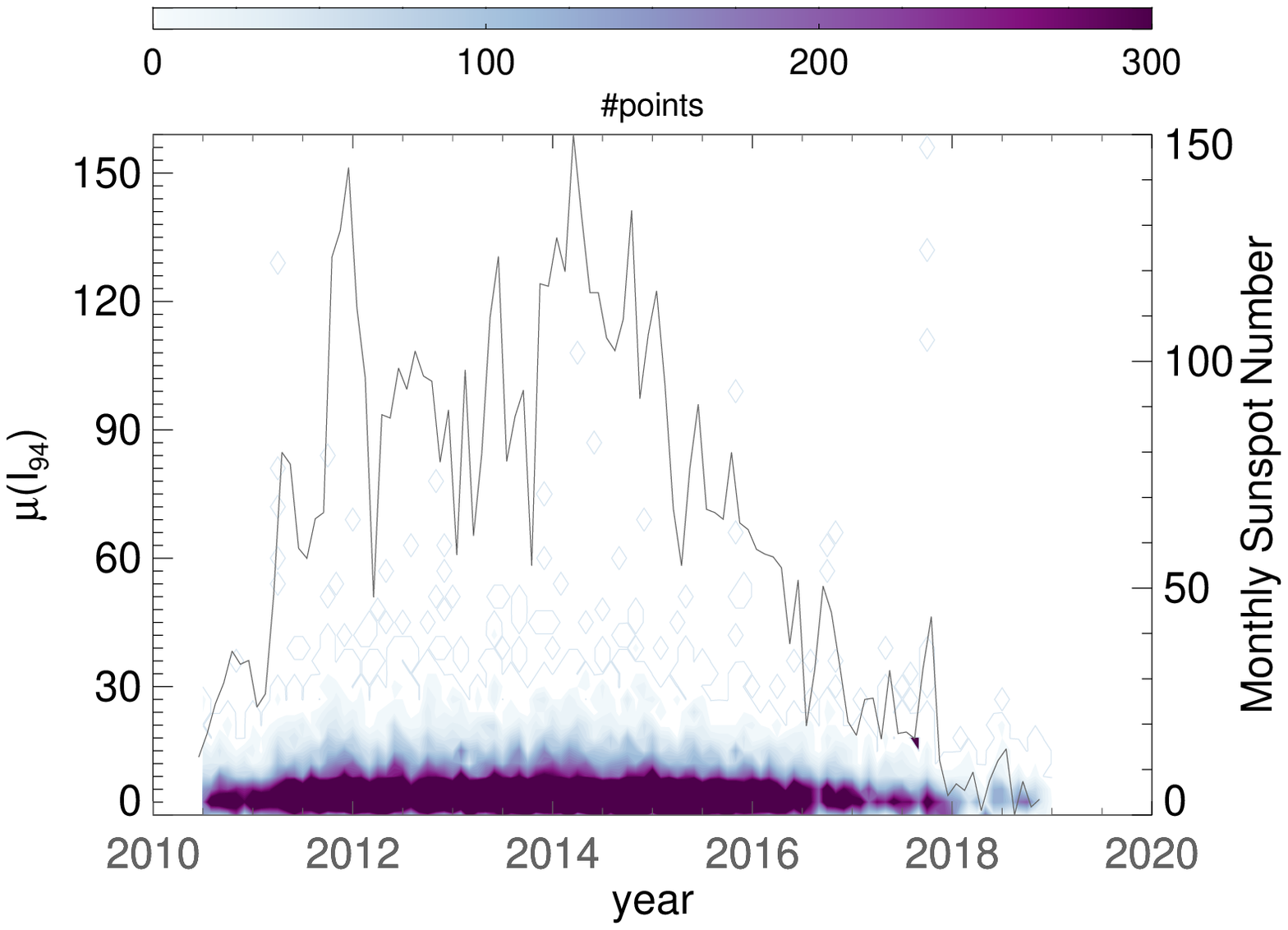}}
\caption{Top Row: Density histograms of the total emission of each AARP for 
171\AA\ (left) and 94\AA\ (right)
as a function of date, left axes.  Also plotted is the monthly mean sunspot number
from \citet{sidc},
right axes.  The density histograms do include the `single-point' 
values as the lowest contour / lightest color.  Bottom Row: Same presentation but
for the mean emission (size-normalized) of the two bandpasses.}
\label{fig:aarp_ssn}
\end{figure}

In Figure~\ref{fig:aarp_ssn} we present density histograms of total emission in
AARPs for two AIA bands over most of Solar Cycle 24, as well as the same
presentation for the mean emission in the same bands.  We focus on 171\,\AA\ that 
is sensitive to plasma at ``quiet'' coronal temperatures and transition
region emission, and thus is often used to identify 
coronal magnetic structures such as coronal loops,
and the 94\,\AA\ band that is sensitive to hot plasma \citep{aia_Lemen}, but
that can contain contributions from cooler lines within the bandpass 
\citep{ODwyer_etal_2010,Warren_etal_2012,DelZanna2013}. 

The ``total'' emission ($\Sigma(I_*)$) distribution tracks nicely with sunspot number, as 
expected, since it is an extensive parameter that scales with the
size of the box, number of pixels, or more physically, the total amount 
of magnetic flux present.  In both bands
we see that the vast majority of AARPs have small total emission, a high
density of points that itself tracks the sunspot number generally.  In both
bands there are also the less-frequent larger-total-emission regions
whose distribution tracks the sunspot number even closer.  None of this
is unexpected.    

What is interesting is that while for most ``bumps'' in total emission,
{\it i.e.} those for which the eye sees a good correlation with a ``bump''
in sunspot number, there is good correspondence between the two 
passbands.  However, there are apparently some large active regions for
which a large total 171\,\AA\ is {\it not} matched with a correspondingly
large total in 94\,\AA\ ({\it e.g.} late 2010, late 2017, early 2018) and 
{\it vice versa} ({\it e.g.} mid 2012, early 2015).  
Thus, while well correlated, the total emission
displayed in these two channels may not be uniquely correlated, which prompts the 
questions ``why or why not?''  One immediate reason is the multi-thermal
nature of both the AIA channels and active regions themselves.  Clearly this extensive
dataset can begin to address the details of active region coronal emission with, {\it e.g.}, 
AR age, magnetic field morphology, and flaring history, and provide constraints
for modeling efforts both with regards to intra-filter expected emission 
and variations with solar cycle.  This is especially true since, as discussed
in sections\,\ref{sec:data_final}, Appendix \ref{sec:appendix_DEM} - the data are ready
for not just hot-plasma isolation \citep{Warren_etal_2012}, but full DEM analysis.

The mean emission tells a slightly different story.
Figure~\ref{fig:aarp_ssn} shows that if there magnetic flux is present
(as defined the HARPs and hence the AARPs), there is emission in 171\,\AA, or 
$\mu(I_{\rm 171}) \gg 0$, essentially.  The same is not quite as true for $\mu(I_{\rm 94})$ where
the majority of AARPs display a low or near-zero mean emission.  This is 
consistent with $\mu(I_{\rm 94})$ having sensitivity to hot coronal
emission, which not all active regions contain (especially flare-quiet ones).  However,
$\mu(I_{\rm 171}) /  \mu(I_{\rm 94})$ may not be perfectly constant with
solar cycle.  
These simple findings are consistent with \citet{Schonfeld_etal_2017}'s
sun-as-a-star investigation showing that the cool / quiet-Sun corona varied little
over the first half of SC24 while the hot component varies strongly with the rise
in solar activity.  A thorough investigation using the AARP database could
provide further physical constraints especially with respect to the 
AR-based contributions to the two components and possibly with regards
to ``terminator'' analysis \citep{LeamonMcIntoshTitle2022}.

\subsection{Active Region Coronal Behavior with Observing Angle}
\label{sec:mu}

\begin{figure}
\centerline{
\includegraphics[width=0.50\textwidth,clip, trim = 0mm 0mm 0mm 0mm, angle=0]{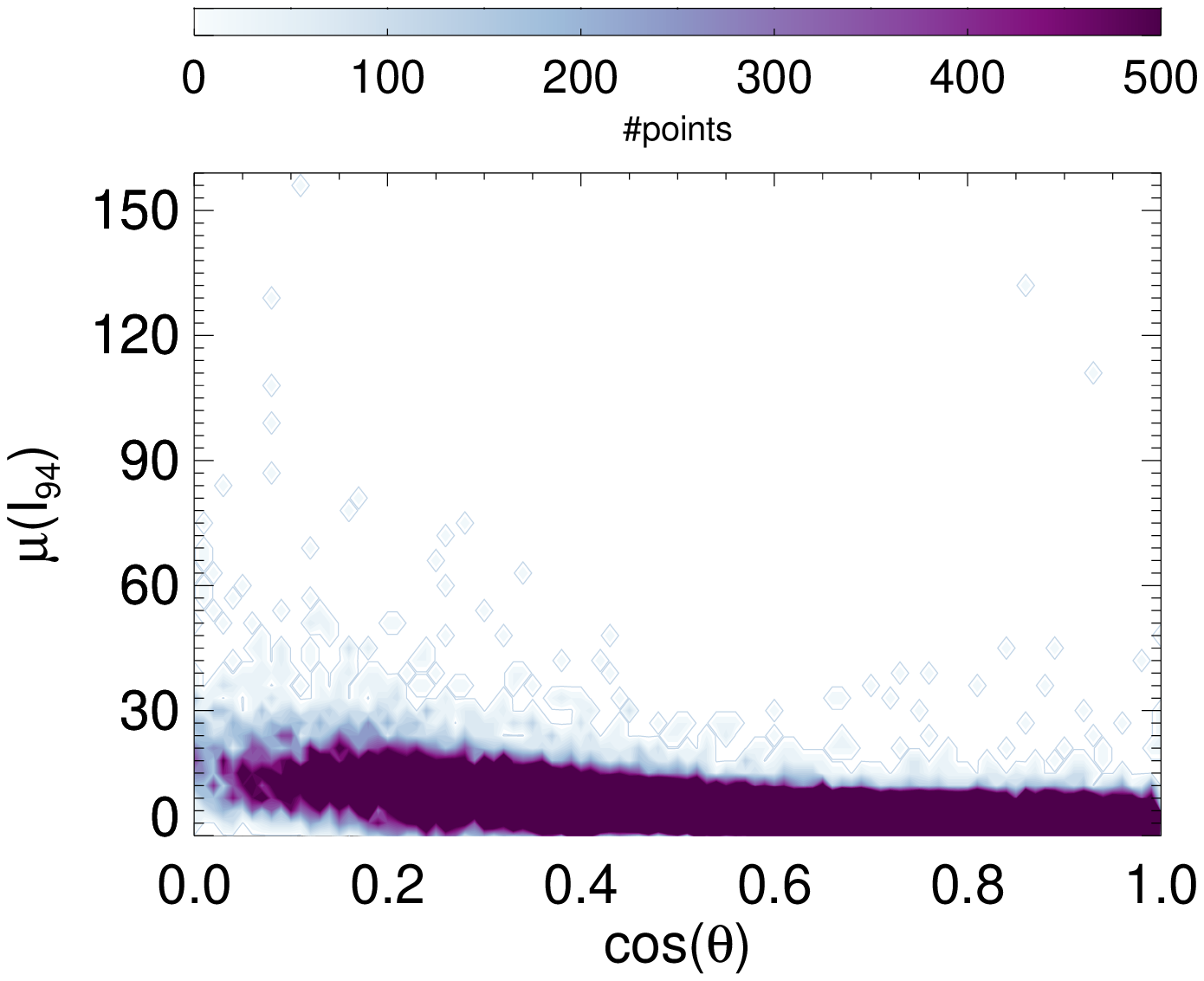}
\includegraphics[width=0.50\textwidth,clip, trim = 0mm 0mm 0mm 0mm, angle=0]{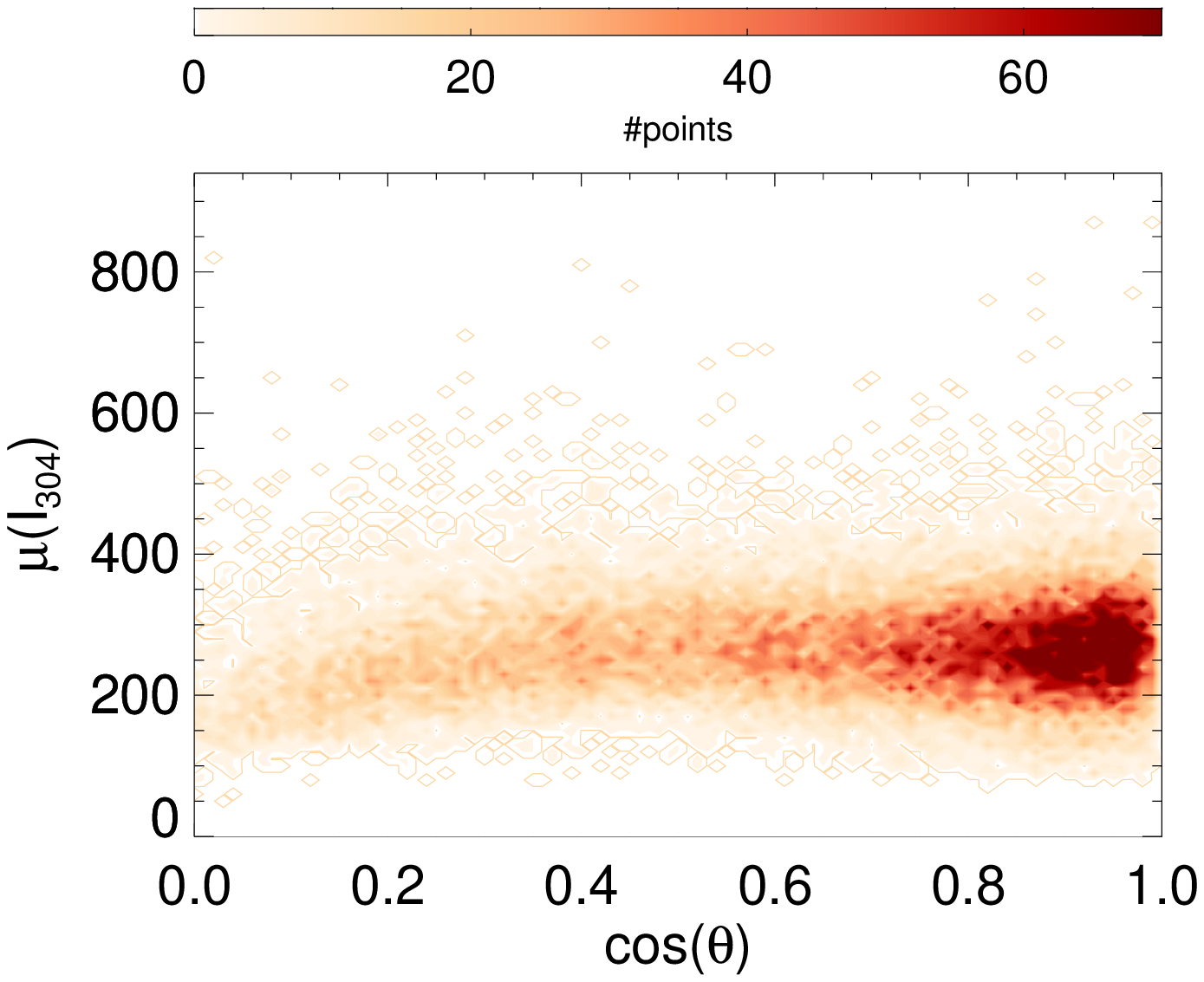}}
\caption{Density histograms of the area-averaged 94\,\AA\ emission (left) 
and 304\,\AA\ emission (right) as a function of observing angle.  Shown
are the results for all AARPs regardless of size or other selection criterion.} 
\label{fig:aarp_mu}
\end{figure}

The coronal emission lines are generally optically thin.  However, 
a few of the AIA bandpasses are centered on, or include, optically
thick emission \citep{Golding_etal_2017}.  Additionally, when densities become large
such as during solar energetic events, some of the normally optically-thin lines in the sampled
bandpasses become optically thick \citep{Thiemann_etal_2018}. 
The differences in behavior can be seen when the
mean emission levels of different lines are presented according to their
location on the disk (Figure~\ref{fig:aarp_mu}).  The 94\,\AA\ channel
generally displays a small mean emission that is flat
with $\mu=\cos(\theta)$ except at the very limb, where the optically-thin properties 
afford a propensity to integrate over more
emitting structures along the line-of-sight.  

The \ion{He}{2} 304\,\AA\ 
emission is optically thick but has challenging radiative
transfer characteristics \citep{Golding_etal_2017}, and thus expected to show a correlation 
with observing angle $\cos(\theta)$ (decreasing toward the limb).
The density distribution of $\mu(I_{\rm 304})$ (Figure~\ref{fig:aarp_mu}, right) does
show a slight turnover beyond $\mu=\cos(\theta) = 0.25\ {\rm or}\ \theta=75^\circ$
but it is extremely slight.  There is no severe or obvious attenuation 
with observing angle, but the behavior of \ion{He}{2} 304\,\AA\ emission 
in active regions may not have yet been studied with large-sample data.
There is the possibility of contamination within the AIA filter by optically-thin 
emission that would mask the \ion{He}{2} 304\,\AA\ behavior, and
it also may be contaminated by the different behavior expected by different
magnetic structures within the AARP fields of view \citep{Mango_etal_1978,Worden_etal_1999}.
The detailed behavior of the corona as inferred from optically-thin dominated {\it vs.} optically-thick
dominated AIA filters for active regions can now, with the AARP database, be studied
in detail with large-sample statistical analysis.

\subsection{Coronal Behavior with Active Region Emergence / Decay}
\label{sec:efr_decay}

The relationships (in the spatial and temporal domains) between coronal
emission and emerging magnetic flux should provide insights into the
energy transfer to the upper atmosphere from the photosphere and below.
Similarly, the relationships between coronal emission and decaying
active regions should elucidate the final transfer of magnetic energy
out of the Sun and the dominant dispersion and dissipative mechanisms
at play.  The AARPs provide the ability to examine these processes by
sampling regions at all sizes and stages of evolution with a curated,
large-sample dataset.

\begin{figure}
\centerline{
\includegraphics[width=0.50\textwidth,clip, trim = 2mm 0mm 6mm 0mm, angle=0]{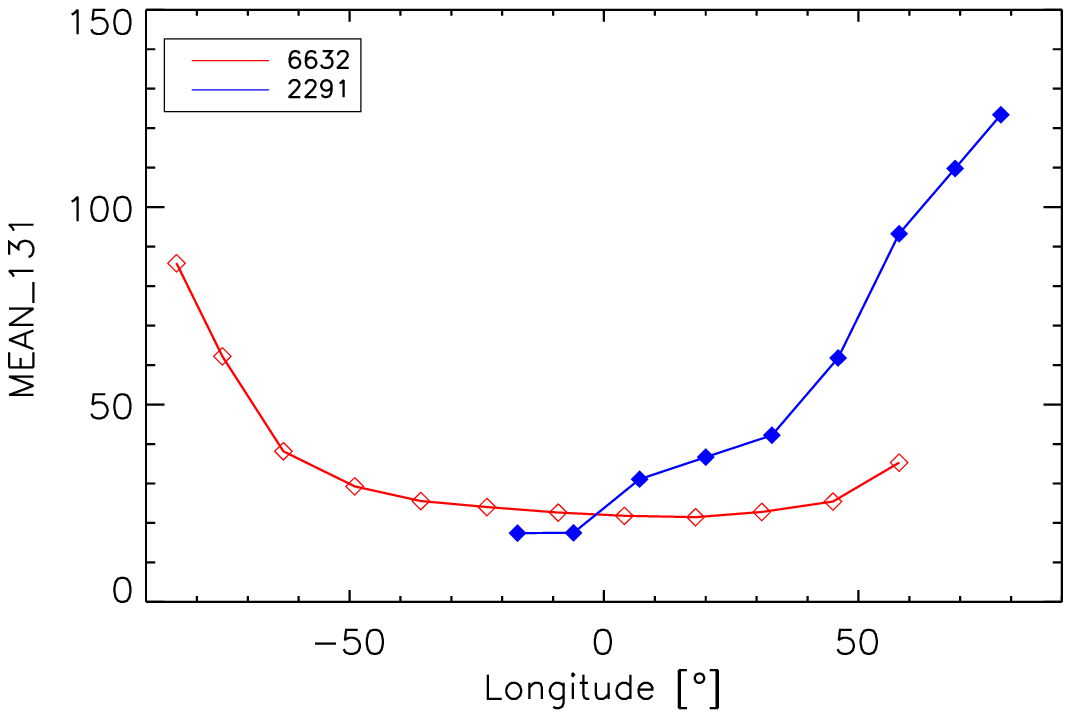}
\includegraphics[width=0.50\textwidth,clip, trim = 2mm 0mm 6mm 0mm, angle=0]{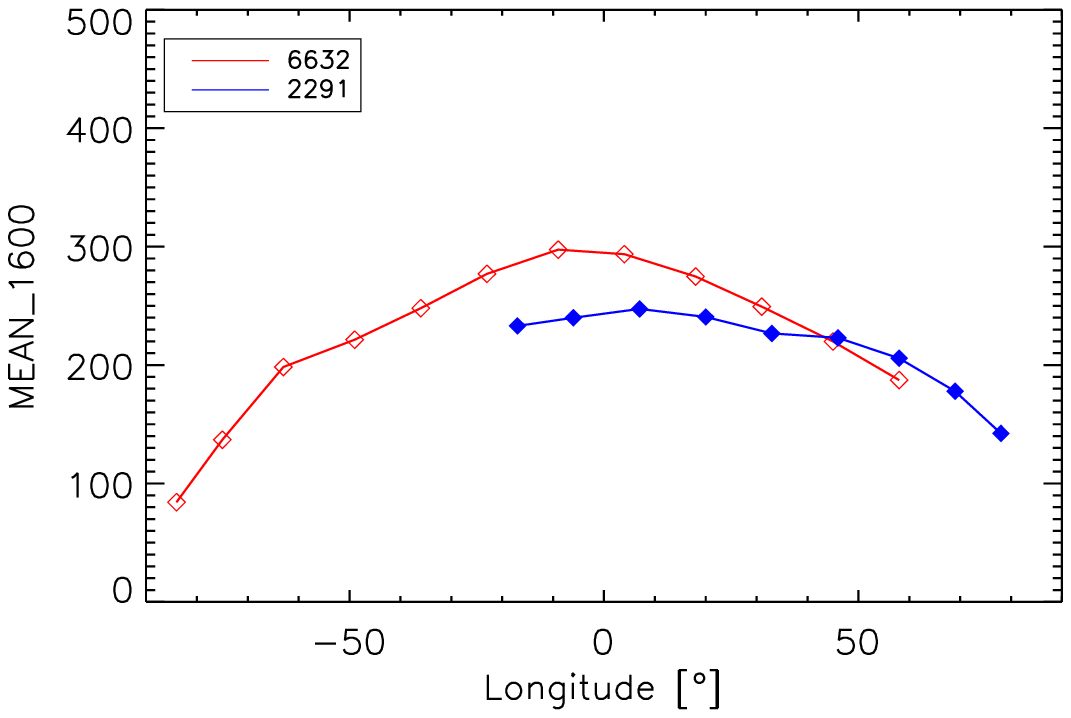}}
\caption{Two HARP/AARP regions and their mean emissions in two AIA filters.  AARP~2291 grew
rapidly starting near disk-center in 2012.12 to become NOAA~AR 11631 and 11632.
AARP~6632 (NOAA~AR 12556, 2016.06) decays during disk passage.  See text for discussion.}
\label{fig:emergence_decay}
\end{figure}

In Figure~\ref{fig:emergence_decay} we show the transition region/flaring plasma sensitive 
$\mu(I_{\rm 131})$ and 
upper-photospheric $\mu(I_{\rm 1600})$ for two AARPs that are in significantly
different stages of evolution as they traverse the disk.  AARP 2291 underwent
rapid growth starting near disk center and eventually became NOAA ARs
11631 and 11632.  It displays a rapid growth in the mean intensity in 
the \ion{Fe}{8,XXI} 131\,\AA~filter emission.  AARP 6632 rotated
on to the disk as NOAA AR 12556 and decayed to plage by disk-center.  Its 
131\,\AA~filter emission decays to essentially background.  The comparison
of the two regions in the $\mu(I_{\rm 1600})$, however, indicate that
(as expected) this emission is optically thick (decreases systematically
toward the limb) and is sensitive to the presence of plage rather than sunspots,
with little difference between young and old plage.

Obviously we simply present a comparison here in order to encourage interest.
We do not, for example, try to perform a larger-sample re-examination of
the relationship between flux transport, flux decay, and active region 
EUV radiance decay \citep{UU_etal_2017} but that is one study the AARPs could
confirm with larger-sample statistics.  Other questions are ready for analysis:
Are there significant differences in coronal characteristics between old and new plage?  
Between emerging regions that will become flare productive and those that will stay quiet?

\subsection{Coronal Behavior with Activity Level}
\label{sec:dir_vs_diff}

The AARP dataset was prepared specifically for a large-sample
investigation into the coronal behavior as related to 
flaring activity.  We defer our study details and results
to Paper II \citep{nci_aia} but here present a preliminary example
of the directions available for study with the AARP dataset.

We note that indeed, the AARPs are not extracted according
to the time of flares, meaning that they are {\it not}
designed for super-posed epoch analysis \citep[{\it e.g.} ][]{Reinard_etal_2010,MasonHoeksema2010,BobraCouvidat2015,Jonas_etal_2018}.
They are, however, extracted for quantitative interpretable analysis
and not flare prediction in \& of itself, as have been many recent
studies using large samples of AIA image data 
\citep[{\it e.g.} ][]{Nishizuka_etal_2017,Jonas_etal_2018,Alipour_etal_2019}.

\begin{figure}[t]
\centerline{
\includegraphics[width=0.50\textwidth,clip, trim = 2mm 0mm 15mm 10mm, angle=0]{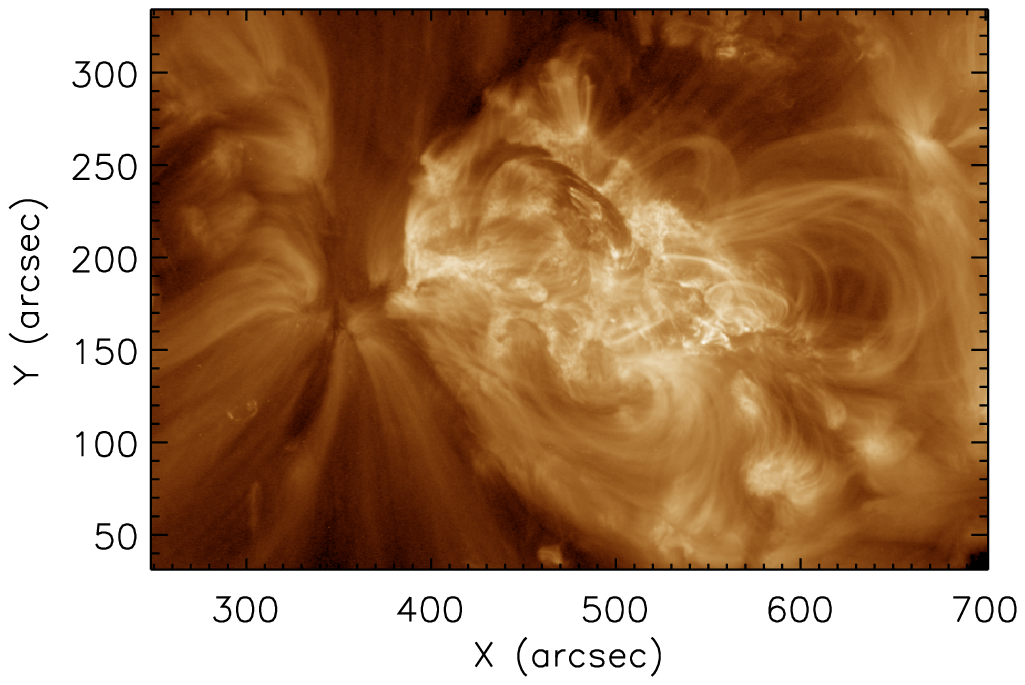}
\includegraphics[width=0.50\textwidth,clip, trim = 2mm 0mm 15mm 10mm, angle=0]{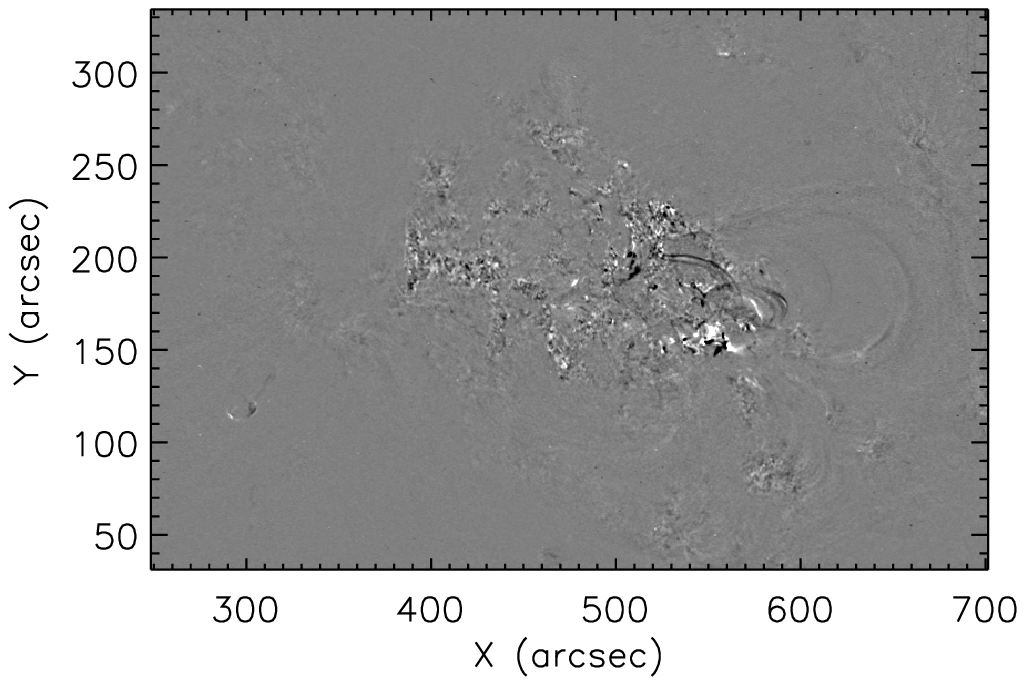}}
\centerline{
\includegraphics[width=0.50\textwidth,clip, trim = 2mm 0mm 15mm 10mm, angle=0]{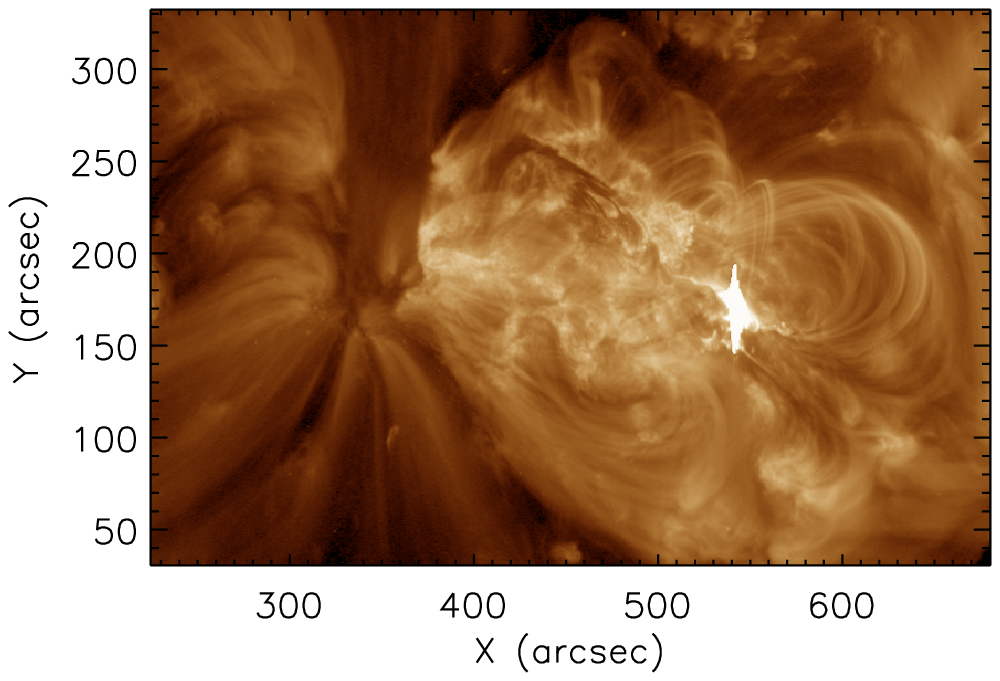}
\includegraphics[width=0.50\textwidth,clip, trim = 2mm 0mm 15mm 10mm, angle=0]{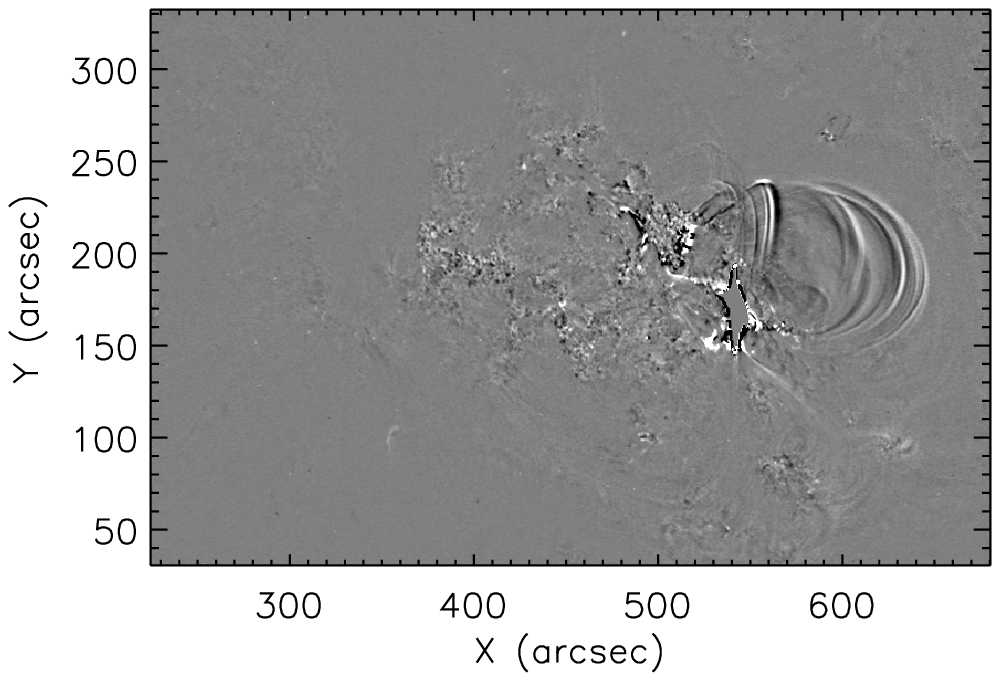}}
\begin{tabular}{|ccccc|cccc|} \hline
 & \multicolumn{4}{c|}{Direct} & \multicolumn{4}{c|}{Running-Difference} \\ 
 & $\mu$ & $\sigma$ & $\varsigma$ & $\kappa$  & $\mu$ & $\sigma$ & $\varsigma$ & $\kappa$ \\ \hline
Quiet & 965.9 & 837.8 & 2.38 & 7.22 & 0.58 & 77.0 & 1.78 & 342.5 \\
Flare & 1022.2 & 1192.6 & 6.87 & 77.8 & -0.87 & 225.2 & -21.2 & 1665.5 \\ \hline
\end{tabular}
\caption{``Direct'' (left) and running-difference (right) images of NOAA AR 11261 
(HMI HARP 750, see Figure~\ref{fig:aiacurves}) on 2011.08.03 showing the differences between a quiet time
(top, 21:46:50-21:45:38~UT) and during a small flare (bottom, 19:46:38-19:45:26~UT).
Of note, the pairs of images are scaled the same, and while the flaring time
does show a small patch of saturation since we avoid the AEC images, 
the pixels around the saturation area have extreme signals in the 
running-difference images as the saturated area changes with time.}
\label{fig:intro_to_params}
\end{figure}

In Figure~\ref{fig:intro_to_params} we show both the ``direct''
(``I$_{193}$'') and running-difference (``$\Delta {\rm I}_{193}$'')
images, in the 193\AA\ band, for NOAA AR 11261 (HMI HARP 750, see
Figure~\ref{fig:aiacurves}) on 2011.08.03.  The quieter example is
from the images centered at 21:48~UT, and the ``active'' example is
taken from images centered at 19:48~UT during a C8.5 flare. The two
``direct'' images look fairly similar, even during a flare, with a
small localized brightening the only easily discernible difference.
The running-difference images look very different.  In the quiet example,
there is a lot of small-scale fluctuation happening even outside the
flare time.  In the flare case, it is true that the saturated-emission area
will show a lack of any dynamics, but do note that as the edges change,
there are strong temporal gradients in this area.  The expanding loops
are also clearly visible in the running difference image.

The images look similar to the eye, as the flare was not very large,
but the images are quantitatively different as summarized by the
moments (mean $\mu$, standard deviation $\sigma$, skew $\varsigma$
and kurtosis $\kappa$) presented in the table associated with
Figure~\ref{fig:intro_to_params}.  In particular, the mean intensity
at the two times is similar for both the direct and running-difference
images (the latter both being fairly close to zero relative to the mean
intensities) but the higher-order moments both for the direct and especially
the running-difference are very different.  Physically this implies enhanced
localized brightening and kinematics on short timescales with little
overall brightness enhancement or mean brightness increase as would
indicate (for example) significant differences in active-region-scale
heating or density between the two time periods.

\section{Summary and Conclusion}
\label{sec:discussion}

We present here the NWRA ``AIA Active Region Patches'' (AARP) Database
\citep{aarp_data}
that well-samples the temporal evolution of solar active regions as
captured in the E/UV with SDO/AIA over the majority of Solar Cycle 24
from 2010/06 -- 2018/12.  The current version of the database includes daily,
7\,hr samples of 13 minutes of images (centered hourly on ``*:48~UT''
from 15:48 -- 21:48~UT) targeting all magnetic patches identified by
SDO/HMI, resulting in a total sample size of 256,976 FITS files containing
in total almost 20 Million individual SDO/AIA images.  We show a teaser
of the analysis that could be performed using this dataset, including
coronal behavior as related to the solar cycle, to the emergence and
decay of solar active regions, as well as center-to-limb variations.

Crucially, the dataset is prepared with attention to quantitative analysis
methodology.  The AARPs dataset preserves the native spatial resolution
of SDO/AIA (i.e. $1.5^{\prime\prime}$ sampled at $0.6^{\prime\prime}$),
in contrast to other studies that downsample the full-disk images to {\it
e.g.}, $512\times512$ \citep[e.g.][]{Galvez:2019} or use solely an active
region's total intensity or maximum intensity in a particular channel
\citep[e.g.][]{Nishizuka_etal_2017}. Keeping the full spatial resolution
allows us to capture small-scale dynamics of the studied active regions
(see e.g.~Figure~\ref{fig:intro_to_params}), which are otherwise lost
due to binning.

However, we note that the dataset is (currently) limited to a selected
time range (to match the NWRA database of HARP-based photospheric vector
field timeseries data \citep{nci_daffs}) although with an extended
field-of-view from the original defining HARP bounding box to capture
the projections of structures that extend in height.  For our initial
purpose of investigating unique coronal and chromospheric characteristics
of flare-imminent active regions, this is a valid approach since both
the short-term changes (over the course of 13 minutes every hour at a
cadence of 72\,s) as well as longer-term trends (7\,hr of evolution per
day per HARP) can be studied.

We stress that the AARP dataset is ready for machine-learning applications
or any other large-sample analysis.  Importantly, it has been validated
for Differential Emission Measure analysis (see Appendix~\ref{sec:appendix_DEM}), which is a data product
to be released in the near future.  We note that this dataset could
be easily expanded to include additional times sampled over each day,
even at a different cadence, such that both forecast-mode and superposed
epoch mode analysis can be performed on every single flare that occurred
since the start of the SDO mission.  Although currently beyond the scope
of this paper and not needed for the analysis presented in Paper II,
such an expansion is quite feasible given the infrastructure NWRA has
now developed.

\begin{acknowledgments}
The authors thank the referee for a thorough reading and insightful feedback
that helped improve the paper.
This work was made possible by funding from AFRL SBIR Phase-I contract FA9453-14-M-0170,
(initial exploration), but primarily by NASA/GI Grant 80NSSC19K0285 with some additional
support from NASA/GI Grant 80NSSC21K0738 and NSF/AGS-ST Grant 2154653. 
We extend our sincere thanks to Sam Freeland for his patience and help with 
the {\tt ssw\_cutout\_service.pro} challenges.
\end{acknowledgments}

\newpage
\appendix \section{The AARP FITS file format}\label{sec:appendix_header}
AARP FITS files contain in total eight extensions. The first extension,
i.e. extension 0 is the primary header and contains no data. A sample
primary header is given in Table \ref{tab:primary_header}. The primary
header is be followed by seven image extensions that contain 11 images
per extension, from 15:48--21:48~UT. An example for an image extension
header is given in Table~\ref{tab:image_header}.

The FITS keyword values are either extracted from the 
original AIA keywords\footnote{\url{http://jsoc.stanford.edu/~jsoc/keywords/AIA/AIA02840_K_AIA-SDO_FITS_Keyword_Document.pdf}} or computed from the WCS coordinates upon 
field-of-view determination and extraction.

\begin{table}[h!]
\footnotesize
\renewcommand{\arraystretch}{0.6}
\begin{tabular}{ l l l l}
\hline \\
\tt SIMPLE & \tt INT & \tt 1 &\\
\tt BITPIX & \tt LONG & \tt 16 & \\
\tt NAXIS & \tt LONG & \tt 0 & \tt NUMBER OF DIMENSIONS\\
\tt EXTEND & \tt INT & \tt 1  & \tt FITS FILE CONTAINS EXTENSIONS\\
\tt INSTRUME & \tt STRING & \tt `AIA' & \tt Instrument name\\
\tt TELESCOP & \tt STRING & \tt `SDO/AIA' &\\
\tt WAVELNTH & \tt LONG & \tt 211  &  \tt [angstrom] Wavelength\\
\tt HARPNUM & \tt LONG & \tt 377 & \tt HMI Active Region Patch number\\
\tt NOAA\_AR & \tt LONG & \tt 11158 & \tt NOAA AR number that best matches this HARP\\
\tt NOAA\_NUM & \tt LONG & \tt 1 & \tt Number of NOAA ARs matching this HARP (0 allowed)\\
\tt NOAA\_ARS & \tt STRING & \tt `11158' & \tt List of NOAA ARs matching this HARP\\
\tt NSAMPS & \tt LONG & \tt 11  & \tt Number of images in each extension\\
\tt NTIMES & \tt LONG & \tt 7  & \tt Total number of times in time series\\
\tt T\_START & \tt STRING & \tt `2011.02.15\_15:48:00\_TAI' & \tt Time of first observation in timeseries\\
\tt T\_STOP & \tt STRING & \tt `2011.02.15\_21:48:00\_TAI' & \tt Time of last observation in time series\\
\tt TREC\_00 & \tt STRING & \tt `2011.02.15\_15:48:00\_TAI' & \tt Extension \#1 time\\
\tt TREC\_01 & \tt STRING & \tt `2011.02.15\_16:48:00\_TAI' & \tt Extension \#2 time\\
\tt TREC\_02 & \tt STRING & \tt `2011.02.15\_17:48:00\_TAI' & \tt Extension \#3 time\\
\tt TREC\_03 & \tt STRING & \tt `2011.02.15\_18:48:00\_TAI' & \tt Extension \#4 time\\
\tt TREC\_04 & \tt STRING & \tt `2011.02.15\_19:48:00\_TAI' & \tt Extension \#5 time\\
\tt TREC\_05 & \tt STRING & \tt `2011.02.15\_20:48:00\_TAI' & \tt Extension \#6 time\\
\tt TREC\_06 & \tt STRING & \tt `2011.02.15\_21:48:00\_TAI' & \tt Extension \#7 time\\
\tt NXMAX & \tt LONG & \tt 692 & \tt Max NAXIS1 over all extensions\\
\tt NYMAX & \tt LONG & \tt 425 & \tt Max NAXIS2 over all extensions\\
\tt NXMIN & \tt LONG & \tt 688 & \tt Min NAXIS1 over all extensions\\
\tt NYMIN & \tt LONG & \tt 421 & \tt Min NAXIS2 over all extensions\\
\tt HISTORY & \tt STRING & \tt `aia\_patches completed...'& \\
\tt END & & & \\
\hline
\end{tabular}
\caption{Sample FITS header of the primary extension (extension 0)
and its keywords. The information contained includes the
wavelength, the HARP number, the associated NOAA active region number,
the central time of each image extension as well as the maximal
dimensions of the image data cubes.}
\label{tab:primary_header}
\end{table}

\begin{table}[h!]
\footnotesize
\renewcommand{\arraystretch}{0.6}
\begin{tabular}{ l l l l}
\hline \\
\tt XTENSION & \tt STRING & \tt `IMAGE' & \tt AIA DATA IMAGE EXTENSION\\
\tt BITPIX & \tt LONG & \tt -32 & \\
\tt NAXIS & \tt LONG & \tt 3 & \tt NUMBER OF DIMENSIONS\\
\tt NAXIS1 & \tt LONG & \tt 692 & \tt X Pixels \\
\tt NAXIS2 & \tt LONG & \tt 421 & \tt Y Pixels\\
\tt NAXIS3 & \tt LONG & \tt 11 & \tt Samples in time\\
\tt T\_REC & \tt STRING & \tt `2011.02.15\_15:48:00\_TAI' & \tt [TAI time]\\
\tt T\_IDX & \tt LONG & \tt 0 & \tt Hours since timeseries start\\
\tt VALID & \tt LONG & \tt 1 & \tt Extension contains valid (1) or no (0) data\\
\tt NIMVALID & \tt LONG & \tt 11 & \tt Valid images out of 11\\
\tt T\_IMG00 & \tt STRING & \tt `2011-02-15T15:42:02Z' & \tt [ISO8601] time for image 00\\
\tt T\_IMG01 & \tt STRING & \tt `2011-02-15T15:43:14Z' & \tt [ISO8601] time for image 01\\
\tt T\_IMG02 & \tt STRING & \tt `2011-02-15T15:44:26Z' & \tt [ISO8601] time for image 02\\
\tt T\_IMG03 & \tt STRING & \tt `2011-02-15T15:45:38Z' & \tt [ISO8601] time for image 03\\
\tt T\_IMG04 & \tt STRING & \tt `2011-02-15T15:46:50Z' & \tt [ISO8601] time for image 04\\
\tt T\_IMG05 & \tt STRING & \tt `2011-02-15T15:48:02Z' & \tt [ISO8601] time for image 05\\
\tt T\_IMG06 & \tt STRING & \tt `2011-02-15T15:49:14Z' & \tt [ISO8601] time for image 06\\
\tt T\_IMG07 & \tt STRING & \tt `2011-02-15T15:50:26Z' & \tt [ISO8601] time for image 07\\
\tt T\_IMG08 & \tt STRING & \tt `2011-02-15T15:51:38Z' & \tt [ISO8601] time for image 08\\
\tt T\_IMG09 & \tt STRING & \tt `2011-02-15T15:52:50Z' & \tt [ISO8601] time for image 09\\
\tt T\_IMG10 & \tt STRING & \tt `2011-02-15T15:54:02Z' & \tt [ISO8601] time for image 10\\
\tt REFHELIO & \tt STRING & \tt `W21S21' & \tt Heliographic pointing string\\
\tt HARPNUM & \tt LONG & \tt 377 & \tt HMI Active Patch number\\
\tt NOAA\_AR & \tt LONG & \tt 11158 & \tt NOAA AR number that best matches this HARP\\
\tt NOAA\_NUM & \tt LONG & \tt 1 & \tt Number of NOAA ARs matching this HARP (0 allowed)\\
\tt NOAA\_ARS & \tt STRING & \tt `11158' & \tt List of NOAA ARs matching this HARP\\
\tt LAT\_FWT & \tt DOUBLE & \tt -20.157000 & \tt [deg] Stonyhurst LAT of flux-weighted cntr pix\\
\tt LON\_FWT & \tt DOUBLE & \tt 20.295200 & \tt [deg] Stonyhurst LON of flux-weighted cntr pix\\
\tt WAVELNTH & \tt LONG & \tt 211 & \tt [angstrom] Wavelength\\
\tt WAVEUNIT & \tt STRING & \tt `angstrom' & \tt Wavelength unit: angstrom\\
\tt T\_REF & \tt STRING & \tt `2011.02.15\_15:48:00\_TAI' & \tt [time] Pointing reference time\\
\tt DATE\_OBS & \tt STRING & \tt `2011-02-15T15:48:00.62' & \tt [ISO8601] Date when observation started\\
\tt T\_OBS & \tt STRING & \tt `2011-02-15T15:48:02.07Z' & \tt [ISO8601] Observation time\\
\tt EXPTIME & \tt DOUBLE & \tt 2.9012420 & \tt [sec] Exposure duration: mean shutter open time\\
\tt EXPSDEV & \tt DOUBLE & \tt 0.00020700000 & \tt [sec] Exposure standard deviation\\
\tt INT\_TIME & \tt DOUBLE & \tt 3.1562500 & \tt [sec] CCD integration duration\\
\tt PIXLUNIT & \tt STRING & \tt `DN' & \tt Pixel intensity unit\\
\tt DN\_GAIN & \tt DOUBLE & \tt 18.300000 & \tt [DN/electron]\\
\tt AECDELAY & \tt LONG & \tt 1536 & \tt AIA\_IMG\_AEC\_DELAY\\
\tt AECTYPE & \tt LONG & \tt 0 & \tt AIA\_IMG\_AEC\_TYPE\\
\tt AECMODE & \tt STRING & \tt `ON' & \tt AIA\_IMG\_AEC\_MODE\\
\tt AIAWVLEN & \tt LONG & \tt 2 & \tt AIA\_IMG\_WAVELENGTH\\
\tt CDELT1 & \tt DOUBLE & \tt 0.60000000 & \tt X Pixel size in CUNIT1 units\\
\tt CDELT2 & \tt DOUBLE & \tt 0.60000000 & \tt Y Pixel size in CUNIT1 units\\
\tt CUNIT1 & \tt STRING & \tt `arcsec' & \tt X pixel units\\
\tt CUNIT2 & \tt STRING & \tt `arcsec' & \tt Y pixel units\\
\tt CRPIX1 & \tt DOUBLE & \tt -201.50000 & \tt [pix] X location of sun center in CCD\\
\tt CRPIX2 & \tt DOUBLE & \tt 622.50000 & \tt [pix] Y location of sun center in CCD\\
\tt CRVAL1 & \tt DOUBLE & \tt 0.0000000 & \tt [arcsec] X origin - center of the solar disk\\
\tt CRVAL2 & \tt DOUBLE & \tt 0.0000000 & \tt [arcsec] y origin - center of the solar disk\\
\tt CROTA2 & \tt DOUBLE & \tt 0.0000000 & \tt [deg] Angle between satellite N and solar N\\
\tt DSUN\_OBS &  \tt DOUBLE & \tt 1.4773886e+11 & \tt [m] Distance from SDO to Sun center\\
\tt RSUN\_OBS & \tt DOUBLE & \tt 971.72022 & \tt [arcsec] of Sun = arcsin(RSUN\_REF/DSUN\_OBS)\\
\tt RSUN\_REF & \tt DOUBLE  &\tt 6.9600000e+08 & \tt [m] Reference radius of Sun\\
\tt INST\_ROT & \tt DOUBLE & \tt 0.056433000 & \tt [deg] Master pointing CCD rotation wrt SDO Z axis\\
\tt IMSCL\_MP & \tt DOUBLE & \tt 0.60075802 & \tt [arcsec/pixel] Master pointing image scale\\
\tt OBS\_VR & \tt DOUBLE & \tt -1972.2185 & \tt [m/s] Speed of observer in radial direction\\
\tt OBS\_VW & \tt DOUBLE & \tt 28105.801 & \tt [m/s] Speed of observer in solar-W direction\\
\tt OBS\_VN & \tt DOUBLE & \tt -1553.5497 & \tt [m/s] Speed of observer in solar-N direction\\
\tt XCEN & \tt DOUBLE & \tt 328.80002 & \tt [arcsec] Ref pixel pointing, arc sec W from sun center\\
\tt YCEN & \tt DOUBLE & \tt -246.90001 & \tt [arcsec] Ref pixel pointing, arc sec N from sun center\\
\tt DEGRAD & \tt STRING & \tt degradation performed \\ & & \tt using calibration \\
& & \tt `version 10' & \tt Degradation correction\\
\tt PCOUNT & \tt LONG & \tt 0 \tt No Group Parameters \\
\tt GCOUNT & \tt LONG & \tt 1 \tt One Data Group \\
\tt CTYPE1 & \tt STRING & 'HPLN-TAN' & \tt Type of X image coordinate axis \\
\tt CTYPE2 & \tt STRING & 'HPLT-TAN' & \tt Type of Y image coordinate axis  \\
\tt END & & &
\end{tabular}
\caption{Sample FITS header of an image extension.}
\label{tab:image_header}
\end{table}

\newpage

\section{Details on Data Acquisition for remote-DRMS sites}
\label{sec:appendix_drms}

In order to produce the AARP dataset, NWRA made extensive use of being
a remote-SUMS/DRMS (Storage Unit Management System/Data Record
Management System) site for the SDO mission.  As such, NWRA can query,
access, and, in some aspects, ``mirror'' the SDO databases in a manner
directly analogous to the host institution (Stanford University) through
the Stanford Joint Science Operations Center (JSOC).  Many institutions
have invested in this capability.  To produce this curated dataset
required numerous steps and considerations that we describe here,
many of which are unique to the SUMS/DRMS system, but 
some were unexpected, hence worth providing to the community.

We did find this approach most tenable ({\it vs.} using the JSOC
on-line interfaces including the {\tt ssw\_cutout\_service.pro},
see Appendix~\ref{sec:appendix_cutout}), in part because we could
simultaneously process multiple HARP-based AARPs that were often present
on the disk, reducing the I/O load and processing time significantly.

Thus, NWRA subscribed to three AIA data series: {\tt aia.lev1} (the ``data 
series''),
{\tt aia.lev1\_euv\_12s}, and {\tt aia.lev1\_uv\_24s} (the ``header series'').
The first queries the actual image data, the latter two query the metadata or
header information for EUV (94, 131, 171, 193, 211, 304, 335\AA) and UV
(1600, 1700\AA) wavelengths respectively; 
the metadata alone, which comes with the subscriptions, is $\approx$100GB.
In order to download and then access
a complete image including its header information within the SUMS/DRMS system, 
both the data and header series must be queried;  
data within JSOC series are referred to by ``prime keys'' (often, but not exclusively,
a reference time such as {\tt T\_REC}) then data are selected further by 
refining keyword searches.  
From servers hosting a remote-SUMS/DRMS system, querying the
SUMS/DRMS database can be done using the command-line {\tt show\_info}
({\url{http://jsoc.stanford.edu/doxygen\_html/group\_show\_info.html}}),
which is callable from within other codes ({\it e.g.} {\tt bash, IDL},
or {\tt Python}) and the output used accordingly.

To construct an AARP data set, the needed image data are downloaded
first then additional queries
are constructed in order to access the correct entries in the SUMS/DRMS for
analysis.  Image data were batch-transferred using the Java Mirroring Daemon (JMD), 
written and implemented by the National Solar Observatories
({\url{http://docs.virtualsolar.org/wiki/jmd}}; {\url{http://vso.tuc.noao.edu/VSO/w/index.php/Main\_Page}}).
Using an example of the 13\,m interval
centered at 15:48:00\,UT on 2011.01.01
for 94\AA, the following steps are taken:

\begin{enumerate}
\item For 13\,m of data at 72\,s cadence (see Section\,\ref{sec:aia_temporaldownselect}),
the ``slot time'' or first target image is 15:42:00.  This somewhat generic timestamp
is not accepted by {\tt aia.lev1} series, so the prime keys are obtained by querying 
the {\tt aia.lev1\_euv\_12} series:\\
{\tt \small \$ show\_info key="T\_REC,T\_OBS" "aia.lev1\_euv\_12s[2011-01-01T15:42:00/13m@72s][? QUALITY=0 ?][? EXPTIME>1.8 and EXPTIME<3.0 ?][? WAVELNTH=94 ?]" \\
>	T\_REC   T\_OBS \\
>	2011-01-01T15:42:02Z    2011-01-01T15:42:03.57Z \\
>	2011-01-01T15:43:14Z    2011-01-01T15:43:15.57Z \\
>	2011-01-01T15:44:26Z    2011-01-01T15:44:27.57Z  \\
... } \\
\noindent
as {\tt T\_REC} is required for the header series ({\tt aia.lev1\_euv\_12})
whereas {\tt T\_OBS} is needed for the data series ({\tt aia.lev1}).
Note that we specify not just wavelength and exposure time (to avoid AEC
data), but the data-quality requirement as well.
In point of fact, we would query {\it e.g.} {\tt 2011.01.01\_15:42:00/13m@72s}
initially and try {\it e.g.} {\tt 2011.01.01\_15:41:48/13m@72s} if fewer
than the expected number of records were present, or if the difference between 
{\tt T\_REC, T\_OBS} was greater than 18\,s.  Note that the wavelength
can be used in the header series as a second prime key, but not in the data series,
so we generally adopted the practice of invoking the keyword request for all.

\item Populate the local {\tt aia.lev1} series with the requested
full-disk data.  To set up the JMD data-transfer request requires the 
{\tt sunum} (storage unit unique identifier that is associated with a
given data series query) and {\tt recnum} (a record number corresponding
to each record in the storage unit) available from the data series using the
prime keys from above: \\
{\tt \small \$ show\_info -rS key="FSN,T\_REC,T\_OBS" "aia.lev1[2011-01-01T15:42:03.57Z/13m@72s][? QUALITY=0 ?][?EXPTIME>1.8 and EXPTIME<3.0?][?WAVELNTH=94?]"\\
>	recnum  sunum   FSN     T\_REC   T\_OBS \\
>	67456313        190140361       18250728        2011-01-01T15:42:04Z    2011-01-01T15:42:03.57Z\\
>	67456361        190140436       18250776        2011-01-01T15:43:16Z    2011-01-01T15:43:15.57Z\\
>	67456409        190140506       18250824        2011-01-01T15:44:28Z    2011-01-01T15:44:27.57Z\\
... \\ }
\noindent
Note here: there can be more than one record per {\tt sunum} (depending on the series),
the returned {\tt T\_REC} differs from that returned from the header series,
and {\tt sunum} information is not available from the header series.

The Filtergram Sequence Number (``FSN'') integers become important later, but can be queried 
and saved at this step.  For each imaging instrument of SDO ({\it e.g. HMI or AIA}) the FSN is a {\it
unique} number for each original image produced by the instrument; we use them to 
validate the correspondence between the two series.  The {\tt aia.lev1} series 
is not designed to accept multiple-entry requests (``{\tt 13m@72s}''), and when
it fails, the queries must be handled individually.

\item We query for the metadata entries with the new {\tt T\_REC} to confirm
the FSN number correspondence: \\
{\tt \small \$ show\_info key="FSN, T\_REC,T\_OBS" "aia.lev1\_euv\_12s[2011.01.01T15:42:04Z/13m@72s][? QUALITY=0 ?][?EXPTIME>1.8 and EXPTIME<3.0?][?WAVELNTH=94?]" \\
>	FSN    T\_REC   T\_OBS  \\
>	18250728	2011-03-01T15:42:02Z    2011-03-01T15:42:03.57Z\\
>	18250776	2011-03-01T15:43:14Z    2011-03-01T15:43:15.57Z\\
>	18250824	2011-03-01T15:44:26Z    2011-03-01T15:44:27.57Z\\
... }

\item Upon data transfer, we query the data series again, for the location of the 
files using the FSN numbers returned by the {\tt aia\_lev1\_euv\_12s}.  Every remote-SUMS/DRMS system will have its own upper-level structure, but
the {\tt sunum} is consistent between them and is integral to the location:\\
{\tt \small \$ show\_info -qP 'aia.lev1[][18250728, 18250776, 18250824, ... ]'\ seg=image\_lev1 \\
> /nwra/SUMS/SUM0/D190140361/S00000/image\_lev1.fits \\
> /nwra/SUMS/SUM0/D190140436/S00000/image\_lev1.fits \\
> /nwra/SUMS/SUM0/D190140506/S00000/image\_lev1.fits \\
... }

\item We proceed to prepare the data by specifying wavelength,
etc. keywords and the file location ({\it e.g.}, {\tt
/nwra/SUMS/SUM0/D190140361/S00000}).  Invoking {\tt drot\_map.pro},
degradation, field of view determination and extraction, coalignment,
{\it etc.}, requires numerous additional keyword-queries through {\tt
show\_info} to the {\tt hmi.Mharp\_720s} series.  Along the processing,
intermediate steps are saved as temporary {\tt IDL} ``idl.sav'' files
for subsequent assembly into the AARP extraction FITS files.

\end{enumerate}

\noindent

The data transfers and referencing were indeed time consuming to automate, as these series are 
constructed with different structure than, {\it e.g.} the HMI series.  However, 
having the full-disk data locally was key while we formulated the spatial extraction
details (see Section\,\ref{sec:aia_spatialdownselect}) and the sensitivity 
corrections (see Section\,\ref{sec:aia_data}), because we were not required to re-request 
full-disk data repeatedly as the respective approaches were refined.

\section{Notes on Using the {\tt SolarSoftware Cutout Service}}
\label{sec:appendix_cutout}

Our initial attempt, over 2014-2015, to construct the AARP timeseries
HARP-congruent dataset relied on JSOC and its on-line interfaces including through
the {\it Solar SoftWare} IDL package.  We used the \citep[SSW][]{solarsoft} tool
{\tt ssw\_cutout\_service.pro} written by Sam Freeland at LMSAL.  This approach is advantageous 
for many studies, as
the AARP-required data can be downloaded and assembled without the
effort and resource allocation necessary to subscribe to the voluminous
DRMS series, instead utilizing LMSAL's own subscription.  Further, the
LMSAL-generated SSW routines are equipped to deal with the subtleties of
AIA data processing, such as how to set the cadence to avoid AEC frames.

We automated various steps required to implement the cutout service. 
Drivers were written specifically to:
\begin{enumerate}

\item Get HMI metadata, using {\tt show\_info} to select desired times, and patch size/location coordinates: \\
{\tt \small SSWIDL>  fovxa = abs(crsize1*cos(crota2)-crsize2*sin(crota2))*cdelt1 \\
    SSWIDL> fovya = abs(crsize2*cos(crota2)+crsize1*sin(crota2))*cdelt1}

\item Parse the AIA metadata, and select start/end times to avoid AEC images, getting specified URLs: \\
{\tt \small SSWIDL> ssw\_jsoc\_time2data, t0, t1, index, urls, /urls\_only,
 wave=wave, ds='aia.lev1\_euv\_12s'}

\item Queue AIA cutout exports corresponding to HMI Active Region Patches
(HARPs), for the given day using {\tt ssw\_cutout\_service}, invoking the option of
an email when the job is complete:

{\tt \small SSWIDL> ssw\_cutout\_service, t0n, t1n, query, stat, ref\_time=tref, fovx=fovxa, fovy=fovya, 
wave=wavelengths, ref\_helio=ref\_helio, instrument='aia',aec=0, 
cadence=cadencestr, description=descr, max\_frames=10000, email=email, /RICE}

\item Wait for the exports to complete, and move exported FITS files into place

\item Read in the FITS files for each HARP/wavelength/time

\item Call {\tt aia\_prep} to perform alignment/adjustments on each data cube: \\
{\tt \small SSWIDL> aia\_prep, indexIn, dataIn, index, data, /cutout, index\_ref=indexIn[refidx]}

\item Save the final processed data cubes to FITS (and IDL .sav) files (one per HARP per day)
\end{enumerate}

All of the above steps were automated, initiated by specifying a date.
The 94, 131, 171, 193, 211, 304, and 335 \AA~filter data can be processed together,
but the different cadence and data series means the 1600 (UV) must
be processed separately.

To produce the full AARP database, however, we came to a few situations
that nudged us to the approach in Appendix~\ref{sec:appendix_drms}.
The main failure point stemmed from the (lack of) robustness of data
retrieval and needed information not being available to easily detect and
validate failures.   Data transfer throttling -- which can occur on either
end of the transfer, and it is not always easy to assess where it occurs
-- plus ``standard-issue network intermittency'' often created problems
accessing records at the query step.  When records then came up absent,
it was not straightforward to determine exactly which were missing.

Additionally, the cutout service proved to be slower than expected,
primarily due to the need to request data by individual HARPs and
individual days, which necessitated repeated full-disk I/O at the host
institution when, for example, there were multiple HARPs on the same day and time step.
We found significant speed up by internally accessing the full-disk data
and performing cutouts {\it once} for all targets per day and time step.

The {\tt ssw\_cutout\_service.pro} works very well for requests that focus on
limited time periods or a small number of targets.  It allows for fully
customized fields-of-view, significant flexibility on cadence and filter selection, and
provides up-to-date pre-processing.  Some challenges that were faced during this
exploration were beyond the purview of {\tt ssw\_cutout\_service.pro}
({\it e.g.} the too-generic JSOC email content).
Thus, while the exploration into using the {\tt ssw\_cutout\_service.pro}
involved significant investment, it also provided good education
and preparation for the switch to relying upon NWRA's remote-DRMS/SUMS
subscription (see Appendix~\ref{sec:appendix_drms}).  

NWRA will share {\tt ssw\_cutout\_service.pro}-related driver codes with the community
upon request.

\section{Pathology}
\label{sec:appendix_pathology}

The AIA Active Region Patch (AARP) Database \citep{aarp_data}
follows the HMI Active Region Patch (HARP) definitions \citep[e.g.][]{hmi_pipe}.  
By default, all HARPs that were identified between \mbox{2010/06 -- 2018/12}
are included.  However, the HARPs include ``active pixel'' maps within
the bounding boxes that are unique to each HARP, even if two (or more)
HARP bounding boxes overlap.  We do not include the masks (such as are
in the {\tt hmi.Mharp\_720s bitmap.fits} segment) as they would only
refer, roughly, to the footpoints of the coronal structures.  Hence,
some AARPs may spatially overlap, or be fully included in one another.
We do not remove or try to disentangle these situations as all, but do
note that generally the ``subsuming'' situation is rare and generally
for which one of the AARPs is very small.

However, we did find pathological HARPs that needed to be excluded,
resulting primarily from periods when the HMI magnetograms were corrupted
in such a way as to flag large sections of limb as ``active pixels''
which then propagated to a HARP definition even though there were
otherwise no magnetic concentrations.  Of note: the following HARP
numbers were removed from the AARP Database: \\
\centerline{{\tt 4276, 4280, 6712, 6713, 6849, 6851, 7207}} \\
\noindent  
In addition, HARP {\tt 4225} was also initially defined with a
corrupted bounding box but in fact there were two NOAA active regions
within, although they took up only a small fraction of the original
bounding box.  We redefined a HARP-appropriate bounding box according
to the active pixel masks ({\tt hmi.Mharp\_720s bitmap.fits} segment,
see \citet{hmi_pipe}), and then extended that FOV as per described
in Section~\ref{sec:aia_spatialdownselect} for inclusion in the AARP
database.

\section{Differential Emission Measure Validation}
\label{sec:appendix_DEM}

The AARP data set is corrected for instrument degradation. In general,
we use the DEM code by \cite{Cheung_etal_2015} to calculate emission
measure, temperature and density maps \citep{nci_dem}. By definition,
this particular DEM code handles instrument degradation by correcting the temperature
response functions returned from {\tt aia\_get\_response.pro} using the image
observation times.

\begin{figure}
\centerline{
\includegraphics[width=0.8\textwidth,clip, trim = 0mm 0mm 0mm 0mm, angle=0]{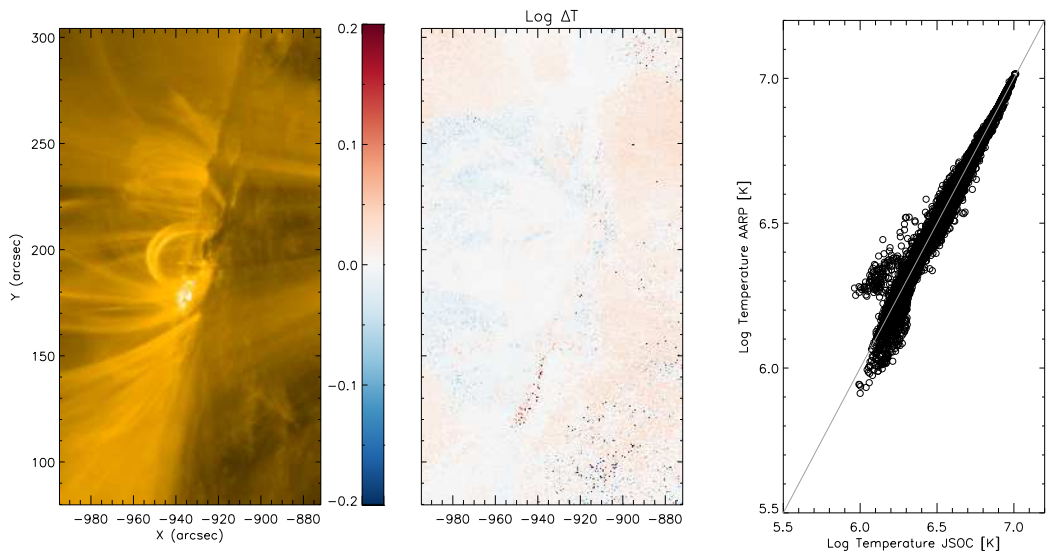}}
\caption{Demonstrating uncertainties in the DEM results using the degradation-corrected AARP
dataset to reconstruct temperature, as compared to using the un-corrected JSOC downloaded
image files (``ground truth'').  Left: 171\,\AA\ image for context; Middle: 
$\log(T_{\text{AARP}}/T_{\text{JSOC}})$ map showing spatial 
differences between the two; Right: Correlation plot between
$T_{\text{AARP}}$ and $T_{\text{JSOC}}$ in $\log$ scale. The gray line
shows the 1:1 correspondence line.}
\label{fig:dem_uncertainty}
\end{figure}

Here, we demonstrate that the DEM calculation using the AARP dataset
as input, and turning ``off'' the degradation correction,
provides similar results as using the original
JSOC output where the degradation has not yet been performed and allowing
the DEM code to make these corrections (Figure~\ref{fig:dem_uncertainty}).
The differences are displayed as both spatial maps and a correlation plot, and
do not exceed the [-0.1,0.1] range in a pixel-by-pixel comparison.  This error range
is in agreement with uncertainties for the sparse solution of \citet{Cheung_etal_2015} when 
compared to a ``ground truth'' ({\it c.f}, their Figures 2 \& 3).

\facilities{SDO (HMI and AIA); GOES (XRS)}
\software{SolarSoft \citep{solarsoft}}

\bibliography{ms}{}
\bibliographystyle{aasjournal}

\end{document}